\documentclass[aps,prb,twocolumn,showpacs,floatfix,twoside,superscriptaddress]{revtex4-2}
\usepackage{amssymb}
\usepackage{graphicx}
\usepackage{amsmath}
\usepackage{amsfonts}
\usepackage{bbold}
\usepackage{esint}
\usepackage{color}
\usepackage[dvipsnames]{xcolor}
\usepackage{float}
\usepackage{verbatim}
\usepackage{subcaption}
\usepackage[percent]{overpic}
\usepackage[colorlinks=true,citecolor=blue,linkcolor=blue,urlcolor=blue]{hyperref}
\usepackage[justification=centerlast,font={small}]{caption}
\usepackage{placeins}

\newcommand{\BT}{\mathcal{T}}

\newcommand{\BP}{\mathcal{P}}

\begin{document}
 
\title{Quantum dynamics of two $XX$ interacting $\BP\BT$--symmetric
non-Hermitian qubits: enhancement of quantum annealing}
\author{Yana Komissarova}
\email{Yana.Komissarova@rub.de}
\author{Mikhail V. Fistul}
\author{Ilya M. Eremin}
\affiliation{Institut f\"ur Theoretische Physik III, Ruhr-Universit\"at Bochum, Bochum 44801, Germany}

\date{\today}

\begin{abstract}
Quantum information platforms enable analog quantum simulations, such as quantum annealing, offering a promising route to solving complex combinatorial optimization problems. Here,  we propose a quantum information architecture based on networks of interacting parity-time ($\mathcal{PT}$)-symmetric non-Hermitian qubits. While the dynamics of individual $\mathcal{PT}$-symmetric qubits have been experimentally demonstrated across multiple platforms—including NV centers, superconducting circuits, and trapped-ion systems—coherent dynamics in interacting systems remain largely unexplored. To address this issue we theoretically investigate stationary and time-dependent Hamiltonians relevant to quantum annealing using a minimal model of two interacting $XX$-coupled $\mathcal{PT}$-symmetric non-Hermitian qubits. We analyze both symmetry-preserving and symmetry-broken regimes and demonstrate that adding even tiny $\mathcal{PT}$-symmetric non-Hermitian terms in the qubits Hamiltonian allows to greatly enhance 
the probability of reaching the ground state after annealing.
\end{abstract}

\maketitle

\section{Introduction}
%


\textit{Optimization problems} are central to a wide range of real-world applications, including theoretical chemistry and materials science \cite{mehta2015collection,kirkpatrick1983optimization}, engineering \cite{sioshansi2017optimization}, logistics and routing \cite{dantzig1959}, and economic modeling \cite{mehta2015collection,islam2012mathematical,LI2026109637}. These problems are typically formulated as the minimization of an objective function over many variables, with the goal of identifying the global minimum of a complex energy landscape \cite{mehta2015collection}. When the variables are discrete—for instance, binary variables equivalent to spin-$1/2$ degrees of freedom—the problem can be mapped onto finding the ground state of a \textit{network of interacting spins} 
\cite{lucas2014ising}.

Determining the ground state of an arbitrary network of interacting spins is, in general, computationally intractable due to the exponential growth of the configuration space with system size. While exact approaches based on exhaustive search exist \cite{quinton2025}, they rapidly become impractical. Consequently, a variety of approximate methods have been developed, including Monte Carlo techniques combined with classical (simulated) annealing \cite{kirkpatrick1983optimization,barzegar2018optimization,heim2015quantum} and, more recently, machine-learning-based approaches \cite{dahrouj2021overview,alcalde2020convolutional}. These methods aim to mitigate the key challenge posed by rugged energy landscapes characterized by numerous local minima separated by large barriers \cite{kirkpatrick1983optimization}.
In many cases, the application of these methods makes it possible to overcome the main obstacle, i.e., the presence of a large number of local minima separated by high barriers in the energy landscape \cite{kirkpatrick1983optimization}. 

Over the past two decades, significant attention has been devoted to artificially engineered networks of interacting two-level systems (qubits), whose dynamics are governed by quantum mechanics at the macroscopic scale \cite{acin2018quantum}. Such systems provide a natural platform for implementing \textit{quantum adiabatic annealing} (QAA) \cite{farhi1998analog,farhi2000quantum,mohammed2025}, a powerful approach to solving optimization problems. QAA relies on the quantum adiabatic theorem \cite{born1928beweis}, which ensures that a system remains in its ground state under sufficiently slow evolution of its Hamiltonian. In practice, one considers a time-dependent Hamiltonian of the form $H(s)=(1-s)H_{\mathrm{in}}+sH_f$, where $s(t)$ varies slowly in time. Here, $H_{\mathrm{in}}$ has an easily preparable ground state, while $H_f$ encodes the solution to the optimization problem. Since its first experimental realizations \cite{johnson2011quantum,boixo2014evidence}, QAA has been implemented across various platforms, including superconducting circuits, photonic systems, cold atoms, and trapped ions, and applied to problems ranging from spin glasses and frustrated systems to traffic optimization and machine learning \cite{boixo2014evidence,king2018observation,bozejko2024optimal,neukart2017,hu2019quantum,yarkoni2022quantum,nath2021review}.In certain cases, it has demonstrated advantages over classical methods \cite{quinton2025,banchi2011nonperturbative}. 

Despite this progress, the broader applicability of QAA is limited by two fundamental challenges: the presence of exponentially small energy gaps and various non-unitary processes, e.g., dissipation. As system size increases, these gaps can become vanishingly small \cite{farhi1998analog,farhi2000quantum}, suppressing quantum tunneling between metastable states, and greatly slowing down the process of QAA.  Moreover, although the robustness of quantum adiabatic algorithms (QAA) against certain non-unitary effects—such as decoherence and noise in time-dependent control parameters of the Hamiltonian \cite{childs2001robustness}—has been demonstrated, it has long been recognized that dissipation arising from specific thermal environments \cite{amin2008thermally,wild2016adiabatic}, as well as coupling to general equilibrium environments \cite{leggett1987dynamics}, can significantly suppress quantum dynamics and thereby degrade annealing efficiency. In addition, numerous attempts \cite{hu2016optimizing,rezakhani2010accuracy,hegde2022genetic} to improve the QAA using the optimization of the time-dependent path $s(t)$, i.e., remaining within the framework of unitary transformations, had a limited success. Therefore, the natural questions arise in this field: is it possible to improve the quantum annealing process by implementing in the QAA Hamiltonian special non-Hermitian terms? and how such non-Hermitian Hamiltonian can be realized in networks of interacting qubits?

To address these questions, we propose and analyze QAA in networks of interacting \textit{$\mathcal{PT}$-symmetric non-Hermitian qubits}. This approach is motivated by recent theoretical and experimental advances in $\mathcal{PT}$-symmetric quantum systems \cite{Wu2023nvu,starkov,gu2022generalized}. It is well established that $\mathcal{PT}$-symmetric non-Hermitian Hamiltonians can exhibit entirely real spectra in the symmetry-preserving phase, or complex-conjugate eigenvalues in the symmetry-broken phase, with the transition occurring at exceptional points \cite{bender1998real,bender1999pt,bender2007}. We also notice that previously a great speed  up ($\ln N$ instead of $N^2$, where $N$ is a number of qubits) of the QAA with local time-dependent non-Hermitian terms has been observed  for the (anti)ferromagnetic Ising integrable model in  \cite{nesterov2013non,nesterov2014non}. 

Experimentally, $\mathcal{PT}$ symmetry was first realized in classical photonic systems via balanced gain and loss \cite{guo2009observation}. More recently, $\mathcal{PT}$-symmetric dynamics have been demonstrated in quantum platforms, including nitrogen-vacancy centers, superconducting circuits, and trapped ions \cite{Wu:2019hsc,naghiloo2019quantum,dogra2021quantum,Kazima,ding2021experimental}. These implementations typically employ dilation schemes, embedding the system into a larger Hilbert space with ancilla qubits and postselection to engineer effective gain and loss.

Building on these developments, we show that introducing a weak $\mathcal{PT}$-symmetric non-Hermitian term can significantly enhance Landau–Zener–Stückelberg (LZS) tunneling, particularly near level crossings. This enhancement mitigates the detrimental effects of small energy gaps and can improve QAA performance in systems where conventional approaches fail. As a minimal example, we analyze QAA in a system of two $XX$-coupled qubits.

The paper is organized as follows. In Sec.~\ref{section:model}, we introduce the general model of a one-dimensional chain of $XX$-coupled $\mathcal{PT}$-symmetric non-Hermitian qubits and define the time-dependent Hamiltonian. In Sec.~\ref{section:energylevels}, we analyze the instantaneous energy spectrum and identify symmetry-preserving and symmetry-broken regimes. In Sec.~\ref{section:effective two-level}, we derive an effective two-level description of the low-energy dynamics. Section~\ref{section:quantum dynamics two-levels} presents numerical and analytical results for the quantum dynamics, including LZS tunneling. In Sec.~\ref{section:QAA original problem}, we return to the full model and analyze the annealing dynamics. Finally, conclusions are given in Sec.~\ref{section:Conclusions}, with technical details provided in Appendix~\ref{section:LZS derivation}.

\section{Chain of $XX$ interacting $\BP\BT$--symmetric non-Hermitian qubits: Model Hamiltonian}\label{section:model}

We start by considering a one-dimensional chain composed of $XX$ interacting $\BP\BT$--symmetric non-Hermitian qubits subjected to effective transverse and longitudinal magnetic fields. In such systems the quantum dynamics of individual qubits is controlled by two parameters, $\Delta$ and $\epsilon$, i.e., a transition amplitude, and a qubit's bias, accordingly. The non-Hermiticity is introduced through a \textit{staggered imaginary} longitudinal magnetic field characterized by a gain (loss) parameter, $\gamma$. All parameters of individual qubits are assumed to be identical. Notice here that $\BP\BT$--symmetric non-Hermitian single qubits have been realized on different experimental platforms \cite{Wu:2019hsc,ding2021experimental,naghiloo2019quantum,dogra2021quantum,Kazima}. 

A simplest way to realize various quantum regimes, and, especially,  the \textit{quantum adiabatic annealing ($QAA$)} process with $\BP\BT$--symmetric non-Hermitian qubits, it is to provide the variation of the transition amplitude, the qubit's bias and the interaction strength, $g$, using a single time-dependent parameter $s(t)$. The Hamiltonian of such a system takes the general form:
\begin{equation}
    \hat H =(1-s)\hat H_{in}+s\hat H_f+\sum_{n}^{N}(-1)^ni\gamma\hat \sigma_n^z,
\label{qa model}
\end{equation}
where $N$ is the number of qubits in the chain. In the \textit{QAA} process \cite{farhi1998analog,farhi2000quantum,johnson2011quantum,boixo2014evidence,king2023quantum} the dimensionless parameter $s$ evolves in time as $s=t/T$ with $0 < t < T$, where $T$ is the duration of quantum annealing. During the \textit{QAA} the Hermitian part of the Hamiltonian (\ref{qa model}) varies from $\hat H_{in}$ to $\hat H_{f}$, while the non-Hermitian part of the Hamiltonian (the third term in the r.h.s. of (\ref{qa model}) ) is keept constant. 

The Hamiltonian $\hat H_{in}$ is chosen in the form of unbiased and non-interacting qubits 
\begin{equation}
    \hat H_{in}=\sum_{n=1}^N\frac{\Delta}{2}\hat\sigma_n^x.
\label{initial hamiltonian}
\end{equation}
The $H_{f}$ contains two terms, the $XX$ coupling between $n$-th and $m$-th qubits with an interaction strength $g(n-m)$, and the qubit's bias, $\epsilon$ 
\begin{equation}
    \hat H_{f}=\frac{1}{2}\sum_{<n,m>}g(n-m)[\hat \sigma_n^x\hat \sigma_m^x+\hat \sigma_n^y \hat \sigma_m^y]+\sum_n\frac{\epsilon}{2}\hat \sigma_n^z \quad, 
\label{final hamiltonian}
\end{equation}
where $\hat \sigma^{x,y,z}$ are the corresponding Pauli matrices.

The parity $\BP$ and the time reversal $\BT$ operators for qubits chains are defined as the operators of the qubits exchange and the complex conjugation, accordingly \cite{STARKOV2023169268,starkov2023schrieffer}:
\begin{equation}
    \hat \BP \hat\sigma_j^r \hat \BP^{-1} = \hat \sigma_{3-j}^r, ~~~~
   \hat \BT i \hat \BT^{-1} = -i. \label{parity_timedef}
\end{equation}
For an even number of qubits the Hamiltonian (\ref{qa model})-(\ref{final hamiltonian}) preserves $\BP\BT$--symmetry for an arbitrary set of parameters $s(t)$, $\Delta$ and $\epsilon$. 

\begin{figure}[H]
\centering
\includegraphics[width=0.8\columnwidth]{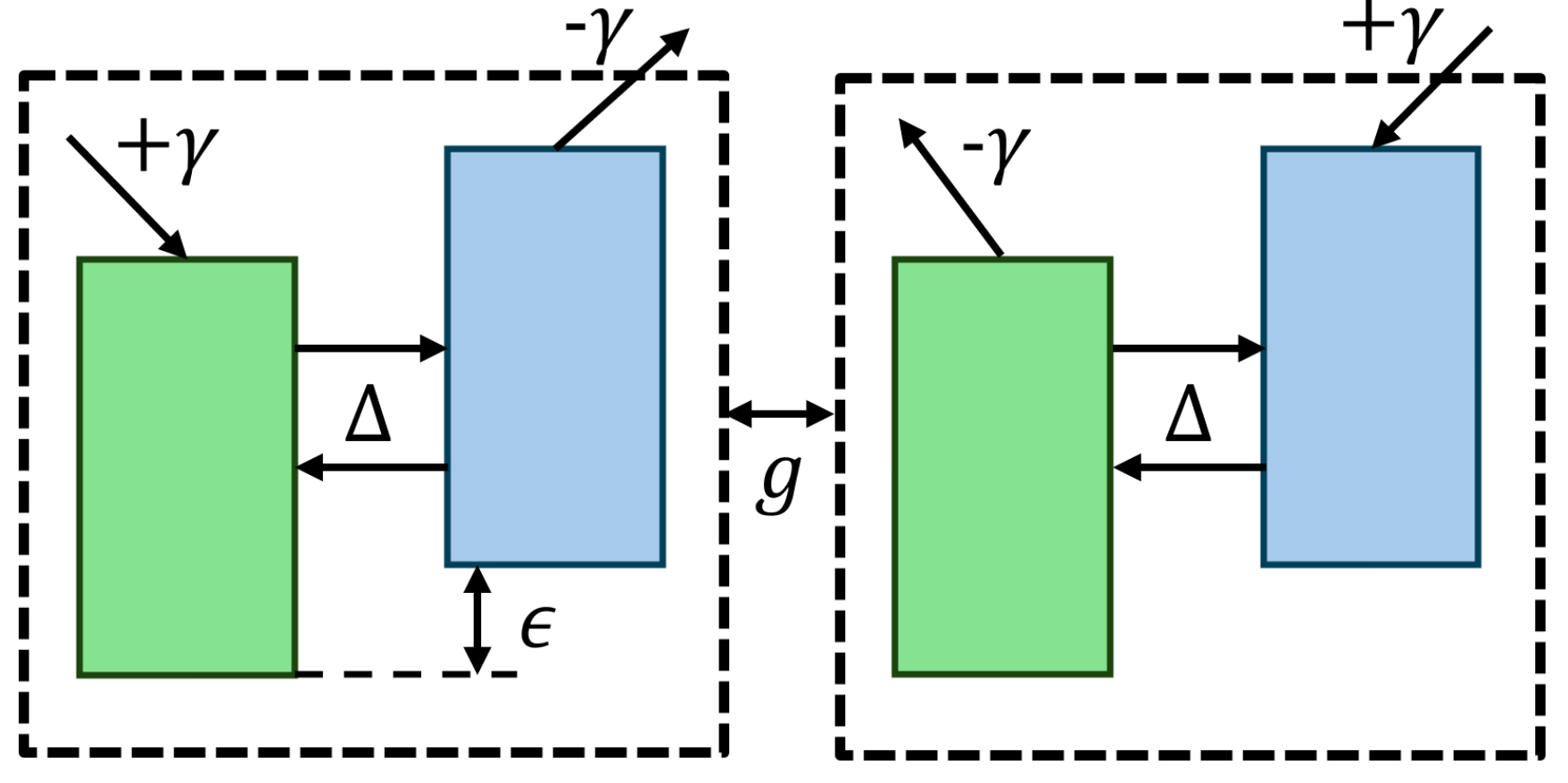}
\caption{Schematic representation of two $\BP\BT$--symmetric  interacting qubits. A transition amplitude between the eigenstates, $\Delta$, a qubit's bias, $\epsilon$, a staggered gain (loss), $\gamma$, and an exchange interaction between the qubits, $g$, are indicated.}
\label{model}
\end{figure}

To analyze the quantum dynamics we take a specific example of \textit{two }$XX$ $\BP\BT$--symmetric non-Hermitian qubits. The model of such a system is presented schematically in  Fig.~\ref{model}.
The Hamiltonian (\ref{qa model})-(\ref{final hamiltonian}) for $N=2$ acquires the form:
\begin{align}
    \hat H = &(1-s) \frac{\Delta}{2}(\hat \sigma_1^x +\hat \sigma_2^x) 
+ s \frac{g}{2} (\hat \sigma_1^x \hat \sigma_2^x +\hat \sigma_1^y \hat \sigma_2^y) +\nonumber \\
&+s \frac{\epsilon}{2}(\hat \sigma_1^z + \hat \sigma_2^z)
+ i\gamma (-\hat \sigma_1^z + \hat \sigma_2^z)
\label{hamiltonian full 2q}
\end{align}
and can be represented as $4\times 4$ matrix:
\begin{equation}
    \hat H = \begin{pmatrix}
s\epsilon & (1-s)\frac{\Delta}{2} & (1-s)\frac{\Delta}{2} & 0 \\
(1-s)\frac{\Delta}{2} & -2i\gamma & sg & (1-s)\frac{\Delta}{2} \\
(1-s)\frac{\Delta}{2} & sg & 2i\gamma & (1-s)\frac{\Delta}{2} \\
0 & (1-s)\frac{\Delta}{2} & (1-s)\frac{\Delta}{2} & -s\epsilon
\end{pmatrix}.
\label{hamilton matrix}
\end{equation}

\section{Two interacting $\BP\BT$--symmetric qubits: instantaneous energy spectrum}\label{section:energylevels} 

\begin{figure*}[!t]
\centering

\begin{minipage}{0.44\textwidth}
  \centering
  \begin{overpic}[width=\linewidth]{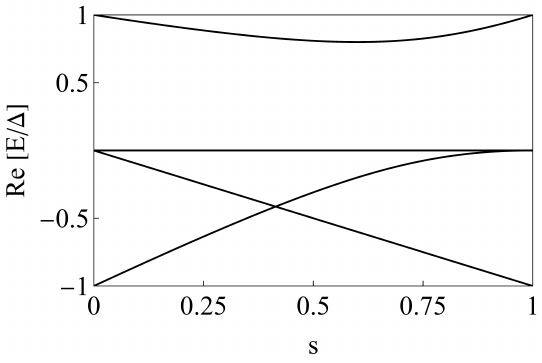}
    \put(20,50){\small\textbf{(a)} $\epsilon/\Delta = 0: \gamma/\Delta = 0$}
  \end{overpic}
\end{minipage}
\hspace{0.04\textwidth}
\begin{minipage}{0.44\textwidth}
  \centering
  \begin{overpic}[width=\linewidth]{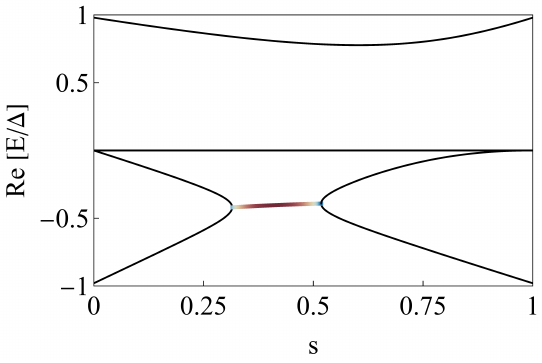}
    \put(20,50){\small\textbf{(b)} $\epsilon/\Delta = 0:\gamma/\Delta = 0.1$}
  \end{overpic}
\end{minipage}

\vspace{2mm}

\begin{minipage}{0.44\textwidth}
  \centering
  \begin{overpic}[width=\linewidth]{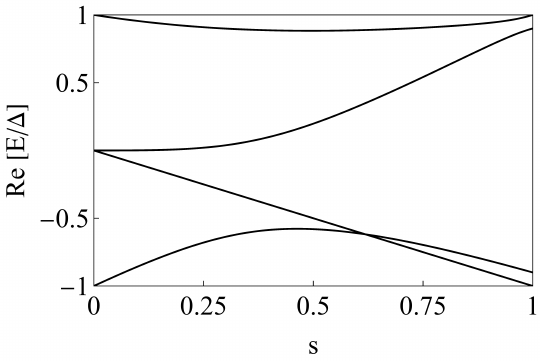}
    \put(20,50){\small\textbf{(c)} $\epsilon/\Delta = 0.9:\gamma/\Delta = 0$}
  \end{overpic}
\end{minipage}
\hspace{0.04\textwidth}
\begin{minipage}{0.44\textwidth}
  \centering
  \begin{overpic}[width=\linewidth]{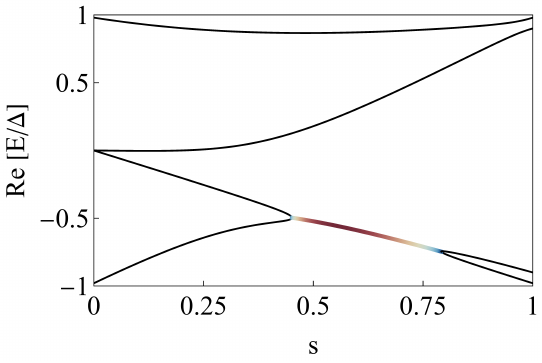}
    \put(20,50){\small\textbf{(d)} $\epsilon/\Delta = 0.9:\gamma/\Delta = 0.1$}
  \end{overpic}
\end{minipage}

\vspace{2mm}
\begin{minipage}{0.44\textwidth}
  \centering
  \begin{overpic}[width=\linewidth]{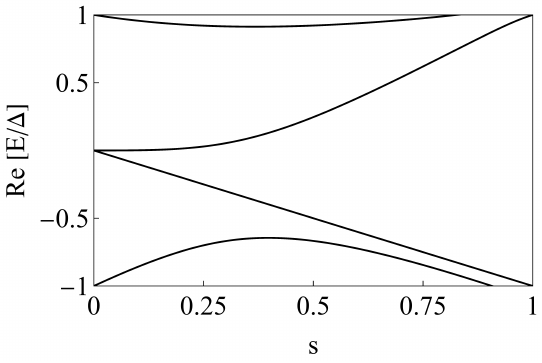}
    \put(20,50){\small\textbf{(e)} $\epsilon/\Delta = 1.1:\gamma/\Delta = 0$}
  \end{overpic}
\end{minipage}
\hspace{0.04\textwidth}
\begin{minipage}{0.44\textwidth}
  \centering
  \begin{overpic}[width=\linewidth]{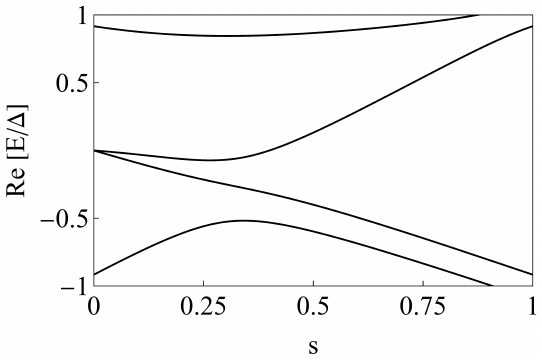}
    \put(20,50){\small\textbf{(f)} $\epsilon/\Delta = 1.1:\gamma/\Delta = 0.2$}
  \end{overpic}
\end{minipage}

\vspace{2mm}

\begin{minipage}{0.44\textwidth}
  \centering
  \begin{overpic}[width=\linewidth]{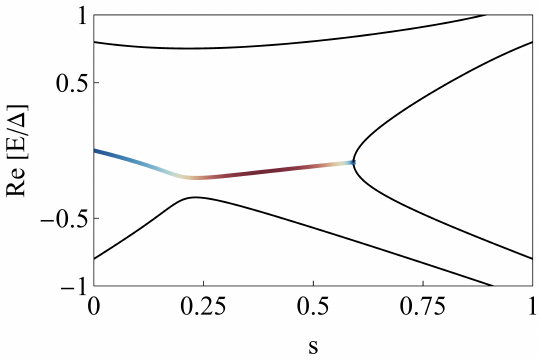}
    \put(20,50){\small\textbf{(g)} $\epsilon/\Delta = 1.1:\gamma/\Delta = 0.3$}
  \end{overpic}
\end{minipage}
\hspace{0.04\textwidth}
\begin{minipage}{0.44\textwidth}
  \centering
  \begin{overpic}[width=\linewidth]{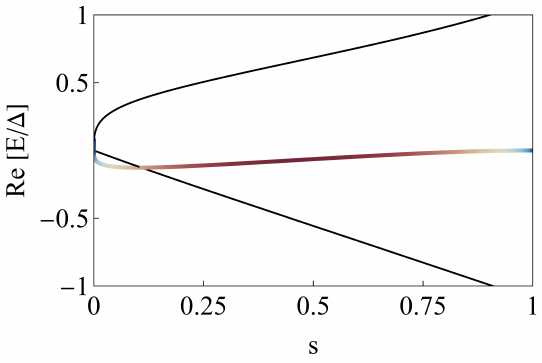}
    \put(18,59){\small\textbf{(h)} $\epsilon/\Delta = 1.1:\gamma/\Delta = 0.5$}
  \end{overpic}
\end{minipage}

\caption{Calculated normalized energy spectrum $\mathrm{Re}[E/\Delta]$ as a function of the external parameter $s$ for $g/\Delta = 1$ and $\epsilon/\Delta = 0$: (a) $\gamma/\Delta = 0$, (b) $\gamma/\Delta = 0.1$; and for $\epsilon/\Delta = 0.9$: (c) $\gamma/\Delta = 0$, (d) $\gamma/\Delta = 0.1$; and for $\epsilon/\Delta = 1.1$: (e) $\gamma/\Delta = 0$, (f) $\gamma/\Delta = 0.2$, (g) $\gamma/\Delta = 0.3$, (h) $\gamma/\Delta = 0.5$. Black curves refer to the real energy eigenvalues, while colored lines correspond to eigenvalues with nonzero imaginary parts. For the colored curves, the real part of the eigenvalue determines the position of the colored line, whereas the imaginary part is encoded in the color gradient, ranging from blue (small imaginary part) to red (large imaginary part).}
\label{spectrum e0}
\end{figure*}

The four eigenvalues of the $\BP\BT$--–symmetric non-Hermitian Hamiltonian~\eqref{hamilton matrix} depending on parameter $s$, i.e., the \textit{instantaneous} energy spectrum $E(s)$, are obtained as solutions of the secular equation: 
\begin{align}
&E^4 + E^2 \left(4\gamma^2 - \Delta^2 - g^2 s^2 - \Delta^2 s^2 - s^2 \epsilon^2 + 2\Delta^2 s\right) + \nonumber \\
&+ E \left(-\Delta^2 g s^3 + 2\Delta^2 g s^2 - \Delta^2 g s\right) 
- 4\gamma^2 s^2 \epsilon^2 + g^2 s^4 \epsilon^2 = 0.
\label{allgem_secular_equation}
\end{align}
Fixing the normalized interaction strength  $g/\Delta=1$ we compute the eigenenergies $E_{1-4}$  for various values of $\gamma/\Delta$ and $\epsilon/\Delta$. It is well known \cite{bender1998real,bender2007making} that the eigenvalues of an arbitrary $\BP\BT$--symmetric non-Hermitian Hamiltonian can take either real or complex conjugate values, and show  in Fig.~\ref{spectrum e0} (for $\epsilon=0$, $\epsilon=0.9$ and $\epsilon=1.1$)
the typical dependencies of the real part of normalized energy levels, $\mathrm{Re}[E(s)/\Delta]$ , with the imaginary part of $E(s)$ indicated by colored areas.  The color gradient 
encodes the magnitude of the imaginary part, with blue denoting small and red-large imaginary components.

In the absence of gain (loss), $\gamma=0$, the Hamiltonian~\eqref{hamiltonian full 2q} becomes Hermitian and all eigenvalues are real values, see Figs.~\ref{spectrum e0}(a,c,e).
Other peculiar property of the energy spectrum $E(s)$ is the intersection of the different energy levels leading to the crossing points characterized by the condition $E_1(s)=E_2(s)$. 

By switching on the gain/loss parameter, $\gamma \neq 0$, in most of regions of $s$ the eigenstates still keep real eigenvalues identifying  $\BP\BT$--symmetry unbroken (preserved)  eigenstates. However, the regions of $s$ with imaginary parts of the eigenvalues also emerge in the spectrum, indicating the onset of the broken $\mathcal{PT}$–-symmetry eigenstates (see, Fig.~\ref{spectrum e0}(b,d,f-h)). 

\begin{figure*}[!t]
\centering

\begin{minipage}{0.44\textwidth}
  \centering
  \begin{overpic}[width=\linewidth]{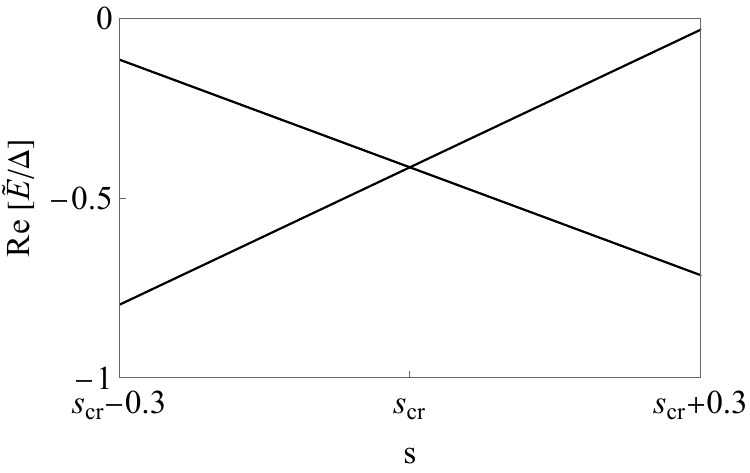}
    \put(32,56){\small\textbf{(a)} $\epsilon/\Delta = 0:\ \gamma/\Delta = 0$}
  \end{overpic}
\end{minipage}
\hspace{0.04\textwidth}
\begin{minipage}{0.44\textwidth}
  \centering
  \begin{overpic}[width=\linewidth]{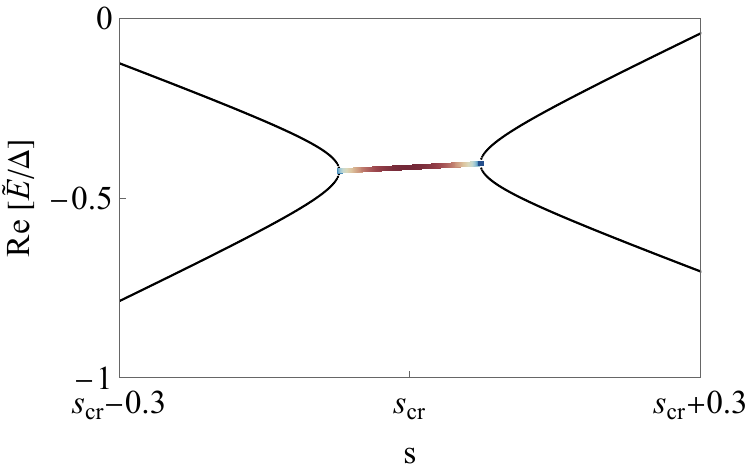}
    \put(32,56){\small\textbf{(b)} $\epsilon/\Delta = 0:\ \gamma/\Delta = 0.1$}
  \end{overpic}
\end{minipage}

\vspace{2mm}

\begin{minipage}{0.44\textwidth}
  \centering
  \begin{overpic}[width=\linewidth]{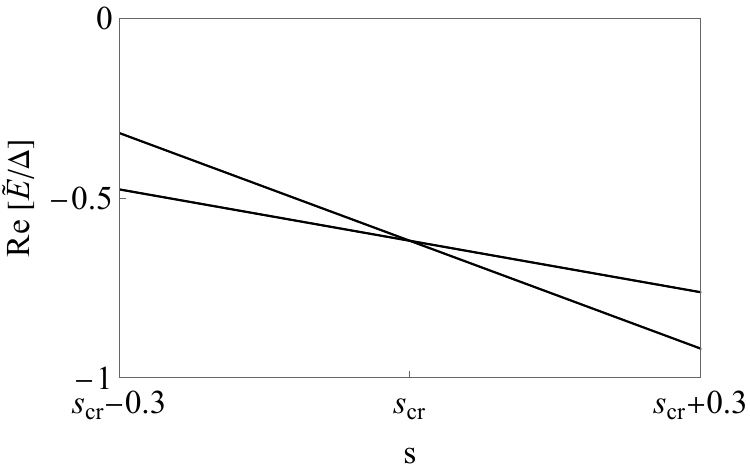}
    \put(32,56){\small\textbf{(c)} $\epsilon/\Delta = 0.9:\ \gamma/\Delta = 0$}
  \end{overpic}
\end{minipage}
\hspace{0.04\textwidth}
\begin{minipage}{0.44\textwidth}
  \centering
  \begin{overpic}[width=\linewidth]{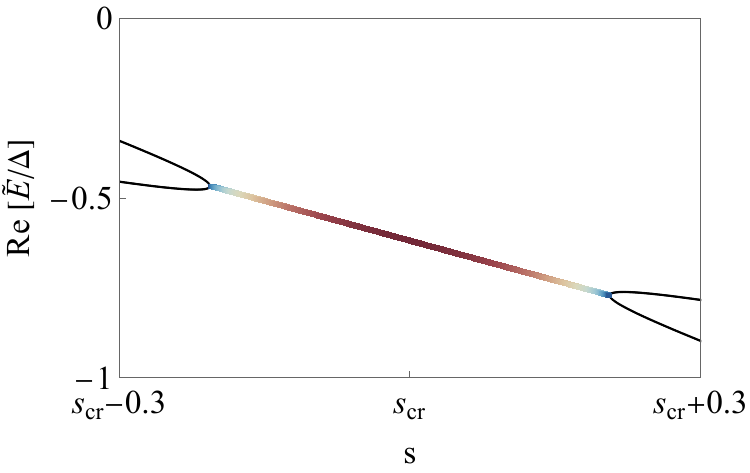}
    \put(32,56){\small\textbf{(d)} $\epsilon/\Delta = 0.9:\ \gamma/\Delta = 0.1$}
  \end{overpic}
\end{minipage}

\caption{Calculated energy spectrum of the effective Hamiltonian~\eqref{Heff gamma} for  (a,b) $\epsilon/\Delta = 0$, $g/\Delta = 1$ and $s_{cr}=0.414$: (a) $\gamma/\Delta = 0$, (b) $\gamma/\Delta = 0.1$; and (c,d) $\epsilon/\Delta = 0.9$, $g/\Delta = 1$ and $s_{cr}=0.619$: (c) $\gamma/\Delta = 0$, (d) $\gamma/\Delta = 0.1$. The color encoding is the same as in Fig.~\ref{spectrum e0}.} \label{spectrum eff}
\end{figure*}

The crossing points transform into the exceptional points (\textit{EP}s) separating the broken $\mathcal{PT}$–-symmetry eigenstates from the unbroken $\BP\BT$--symmetry ones. The second-order \textit{EP}s are characterized by the coalescence of two egenvalues, i.e., $\mathrm{Re}[E_1(s)]=\mathrm{Re}[E_2(s)]$ and  $\mathrm{Im}[E_1(s)]=-\mathrm{Im}[E_2(s)]=0$ . We identify the second-order \textit{EP}s in Fig.~\ref{spectrum e0}(b,d,g,h).
As $\epsilon/\Delta <1$ the region of $s$, where the broken $\mathcal{PT}$–-symmetry eigenstates are realized, expands with an increase of $\gamma$, and the corresponding $EP$s shift to the boundary values of $s=0$ and $s=1$. Notice also that as $\epsilon/\Delta$ is close to one, the two lowest eigenvalues almost coincide for $s \simeq 1$. 


In the regime of $\epsilon/\Delta > 1$, the behavior of the spectrum changes qualitatively. As shown in Fig.~\ref{spectrum e0}(e-h), the crossing of the two lowest energy levels no longer occurs in the Hermitian case. Instead, upon increasing the gain/loss parameter, the second and third energy eigenvalues approach each other and eventually coalesce, marking the onset of a $\BP\BT$--symmetry–broken phase (see Fig.~\ref{spectrum e0}(f,g)). With a further increase of the normalized gain/loss parameter, the real parts of the 
complex-conjugate eigenvalues approach and eventually coincide with the ground-state energy level. Beyond this point, a new pair of complex-conjugate eigenvalues emerges, as illustrated in Fig.~\ref{spectrum e0}(h), indicating the formation of a new $\BP\BT$--broken region.

\section{Two Interacting $\BP\BT$--symmetric non-Hermitian qubits: effective $\BP\BT$--symmetric two-level system}\label{section:effective two-level}
The Figs.~\ref{spectrum e0}(b),(d) clearly demonstrate that in the presence of  gain/loss, $\gamma \neq 0$, the dependencies of upper energy levels on $s$ almost do not change. In contrast, the two lower energy levels are strongly modified in the presence of even weak non-Hermiticity, $\gamma/\Delta<1$. Therefore, it is plausible to assume that the quantum dynamics of lowest energy states of the original $4\times 4$ Hamiltonian \eqref{hamiltonian full 2q} can be reduced to an effective $\BP\BT$--symmetric two-level system. 

The quantum dynamics of such a $\BP\BT$--symmetric two-level system is determined by the effective $\BP\BT$--symmetric non-Hermitian 
$2\times 2$ Hamiltonian obtained as follows. For a weak gain/loss, $\gamma/\Delta<1$, significant changes in the energy spectrum occur near the crossing point of two lowest energy levels (see, Fig.~\ref{spectrum e0}(a),(b), and, first, we obtain explicitly the crossing point values, $s_{cr}$ and $E(s_{cr})$. 
To do this, we simplify the secular equation~(\ref{allgem_secular_equation}) for the case of $\gamma = 0$ as 
\begin{align}
&E^4 + E^2 \left( - \Delta^2 - g^2 s^2 - \Delta^2 s^2 - s^2 \epsilon^2 + 2\Delta^2 s \right) \nonumber \\
&+ E \left(-\Delta^2 g s^3 + 2\Delta^2 g s^2 - \Delta^2 g s \right) + g^2 s^4 \epsilon^2 = 0
\label{secular_equation_gamma_zero}
\end{align}
The value of the parameter $s_{cr}$ at which the lowest energy levels intersect depends on the qubit's bias $\epsilon$, and reads as 
\begin{equation}
s_{cr} =\frac{\sqrt{2} \sqrt{\Delta ^2 g^2-\Delta ^2 \epsilon ^2}-\Delta ^2}{-\Delta ^2+2 g^2-2 \epsilon ^2}
\label{s_cr}
\end{equation}
The corresponding crossing point energy $E(s_{cr})$ is obtained as $E(s_{cr})=-gs_{cr}$.

Next, defining the effective energy $\tilde E=E-E(s_{cr})$ and the controlled parameter $\tilde s=s-s_{cr}$ we expand Eq. (\ref{allgem_secular_equation}) over small values of $\tilde E/\Delta$ and $\tilde s$ up to the second order, and obtain the quadratic secular equation
\begin{equation}
(\tilde E+g\tilde s)(\tilde E+w\tilde s)+\gamma^2(4g^2s^2_{cr}-4s^2_{cr}\epsilon^2)=0, 
\label{sec eq gamma}
\end{equation}
where $w(\epsilon)$ is expressed as
\begin{align}
w(\epsilon) =& \frac{g s_{cr} \left(2 \Delta ^2 (s_{cr}-1)-g^2 s_{cr}+5 s_{cr} \epsilon ^2\right) }{\left(-5 g^2 s_{cr}^2+s_{cr}^2 \epsilon ^2+\Delta ^2 (s_{cr}-1)^2\right)^2}\nonumber \\
&\times\left(-\Delta ^2+5 g^2 s_{cr}^2-\Delta ^2 s_{cr}^2-s_{cr}^2 \epsilon ^2+2 \Delta ^2 s_{cr}\right).
\end{align}
In the absence of qubit's bias $\epsilon=0$ and $g/\Delta=1$, we get $w/\Delta=-1.276$ and $s_{cr}=-1+\sqrt{2}\approx0.414$. In this case the dependencies of instantaneous energy spectrum $\tilde E (s)$ on the parameter $s$ are presented in Figs.~\ref{spectrum eff}(a,b), and one can see a great similarity with the lowest energy curves of the original Hamiltonian ~\eqref{hamiltonian full 2q} (compare, Figs.~\ref{spectrum eff}(a,b) and Figs.~\ref{spectrum e0}(a,b)).

For the biased qubits as $\epsilon/\Delta=0.9$ and $g/\Delta=1$, we get $w/\Delta=0.477$ and $s_{cr}=0.619$, and corresponding dependencies of $\tilde E(s)$ are presented in Fig.~\ref{spectrum eff}(c,d). Similarly to the unbiased qubits case the energy spectrum of the

$\BP\BT$--symmetric effective two-level system demonstrates the same form as the lowest energy spectrum of the original Hamiltonian~\eqref{hamiltonian full 2q} 
(compare, Figs.~\ref{spectrum eff}(c,d) and Figs.~\ref{spectrum e0}(c,d)). Notice here that the dependencies of two energy levels on the parameter $s$ show positive/negative slope for unbiased/biased system of qubits. 

Using the quadratic secular equation \eqref{sec eq gamma} we write the effective $2\times 2$ Hamiltonian of the $\BP\BT$--symmetric two-level system in the following form:
\begin{equation}
\hat H_{eff}=\left(
\begin{array}{cc}
 -g \tilde s & i\ell  \\
 i\ell  & -w \tilde s \\
\end{array}
\right)
\label{Heff gamma}
\end{equation}
with $l=2\gamma s_{cr}\sqrt{g^2-\epsilon^2}$. The matrix (\ref{Heff gamma}) is written in terms of the basis states of the Pauli operator 
$\hat \sigma_x$. 


\begin{figure*}[!t]
\centering

\begin{minipage}{0.3\textwidth}
  \centering
  \begin{overpic}[width=\linewidth]{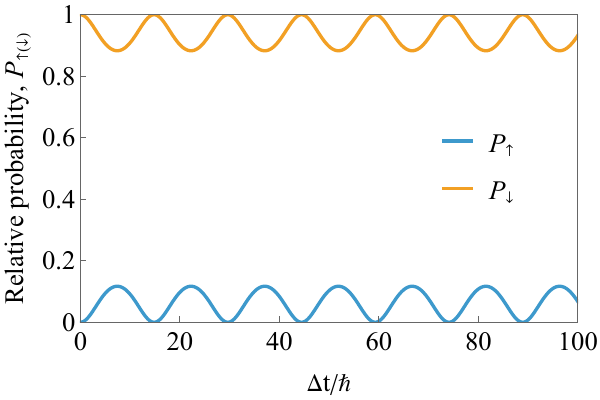}
    \put(20,50){\small\textbf{(a)} $\tilde s_0 = \pm 0.2$}
  \end{overpic}
\end{minipage}
\hspace{0.02\textwidth}
\begin{minipage}{0.3\textwidth}
  \centering
  \begin{overpic}[width=\linewidth]{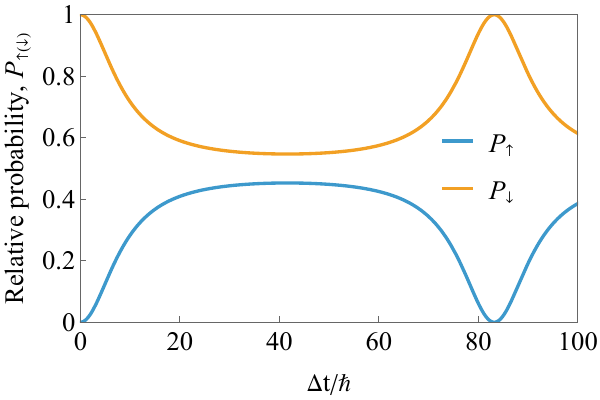}
    \put(20,57){\small\textbf{(b)} $\tilde s_0 = \pm 0.08$}
  \end{overpic}
\end{minipage}
\hspace{0.02\textwidth}
\begin{minipage}{0.3\textwidth}
  \centering
  \begin{overpic}[width=\linewidth]{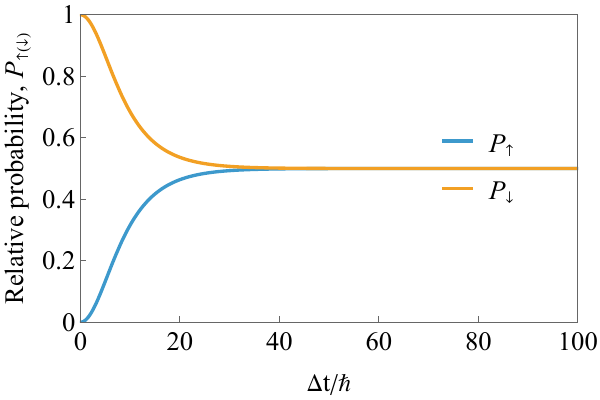}
    \put(20,57){\small\textbf{(c)} $\tilde s_0 = 0$}
  \end{overpic}
\end{minipage}
\caption{Calculated time evolution of the relative population probabilities $P_{\uparrow(\downarrow)}(t)$ of the basis states, $|\uparrow\rangle$ and $|\downarrow \rangle$, 
for fixed values of $\tilde s_0$: (a) $\tilde s_0 = \pm 0.2$, (b) $\tilde s_0 = \pm 0.08$, (c) $\tilde s_0 = 0$. Other parameters were chosen as $\epsilon/\Delta = 0$, $g/\Delta = 1$ and $\gamma/\Delta = 0.1$.}
\label{relative prob plots}
\end{figure*}

\section{Quantum dynamics of the effective $\BP\BT$--symmetric two-level system }\label{section:quantum dynamics two-levels}

The quantum dynamics of the effective $\BP\BT$--symmetric two-level system is governed by the Scr\"odinger equation with the $\BP\BT$--symmetric non-Hermitian Hamiltonian (\ref{Heff gamma}), which, in turn, is controlled by the parameter $\tilde s(t)$, i.e.,
\begin{align}
    i\hbar \frac{d\Psi (t)}{dt}=\hat H_{eff} [\tilde s(t)] \Psi (t),
\label{SEQq-general}
\end{align}
where $\Psi(t)$ is a wave function of the system. The $\Psi(t)$ is represented as
\begin{equation}
\begin{split}
 \Psi(t)=\begin{pmatrix}
\psi_\uparrow (t) \\
\psi_\downarrow (t)
\end{pmatrix}=c_\uparrow (t)\begin{pmatrix}
1 \\
0
\end{pmatrix}+c_\downarrow (t) \begin{pmatrix}
0 \\
1
\end{pmatrix} = ~~~~~~~~~~~~~~~~~~~~~~~~~~ \\
=c_\uparrow (t)|\uparrow\rangle+c_2(t)|\downarrow \rangle.~~~~~~~~~~~~~~~~~~~~~~~~~~~~~~~~
\end{split}
\end{equation}
The Eq. \eqref{SEQq-general} has to be accompanied by the initial condition which we choose to be the ground state of the Hermitian part of (\ref{Heff gamma}) for $s =0$ ($\tilde s=-s_{cr}$), i.e.,
\begin{equation}
    \Psi(s=0)=\begin{pmatrix}
0 \\
1
\end{pmatrix}=|\downarrow\rangle.
\label{psi init eff}
\end{equation}

Since in non-Hermitian systems the time evolution is non-unitary and, therefore, the total probability is not conserved, the quantum dynamics of two-level $\BP\BT$--symmetric non-Hermitian systems has to be characterized by the relevant time-dependent population probabilities $P_{\uparrow(\downarrow)}(t)$ defined as
\begin{equation}
    P_{\uparrow(\downarrow)}(t)=\frac{|\psi_{\uparrow(\downarrow)}|^2}{|\psi_\uparrow|^2 +
|\psi_\downarrow|^2}.
\label{Probability def}
\end{equation}

Next, we numerically calculate $P_{\uparrow(\downarrow)}(t)$ in two regimes determined by the time dependence of the parameter $\tilde s(t)$: $\tilde s=const$  and $\tilde s=t/T-s_{cr}$, where the time $t$ varies between $0$ and $T$. The latter regime provides \textit{QAA } procedure. 

\subsection{Time-independent effective Hamiltonian: $\tilde s=const$}
As the parameter $\tilde s$ is fixed to a specific point $\tilde s_0$ the Hamiltonian~\eqref{Heff gamma} becomes time-independent one, and one can use the time evolution operator to numerically obtain the time-dependent wave function $\Psi(t)$ as following: $\Psi(t)=e^{-it \hat H_{eff}(\tilde s_{0})/\hbar}\Psi(t=0)$, and the corresponding  population probabilities $P_{\uparrow(\downarrow)}(t)$ (Eq.~\eqref{Probability def}).  

\begin{figure*}[!t]
\centering

\begin{minipage}{0.44\textwidth}
  \centering
  \begin{overpic}[width=\linewidth]{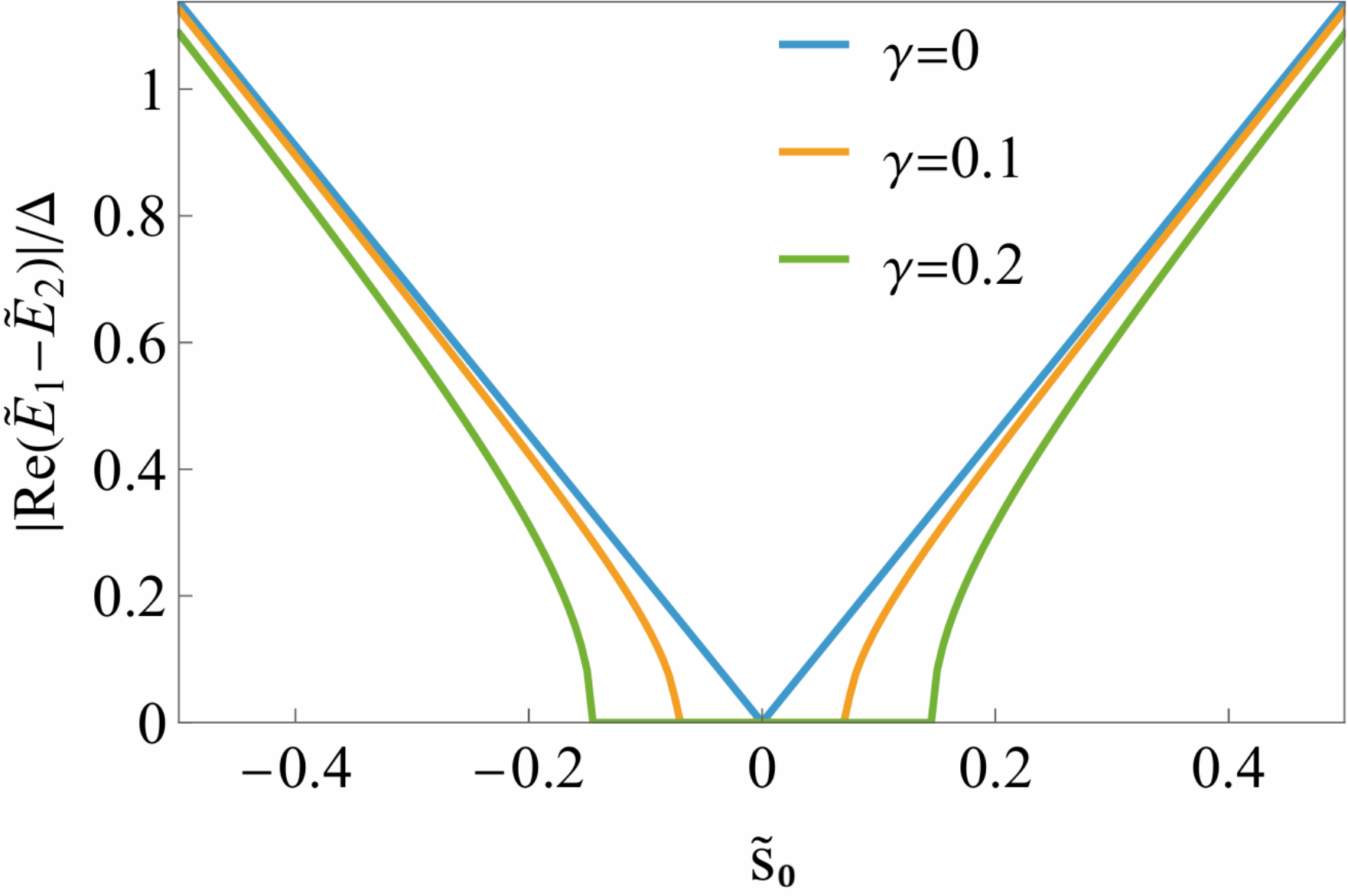}
    \put(23,60){\small\textbf{(a)}}
  \end{overpic}
\end{minipage}
\hspace{0.04\textwidth}
\begin{minipage}{0.44\textwidth}
  \centering
  \begin{overpic}[width=\linewidth]{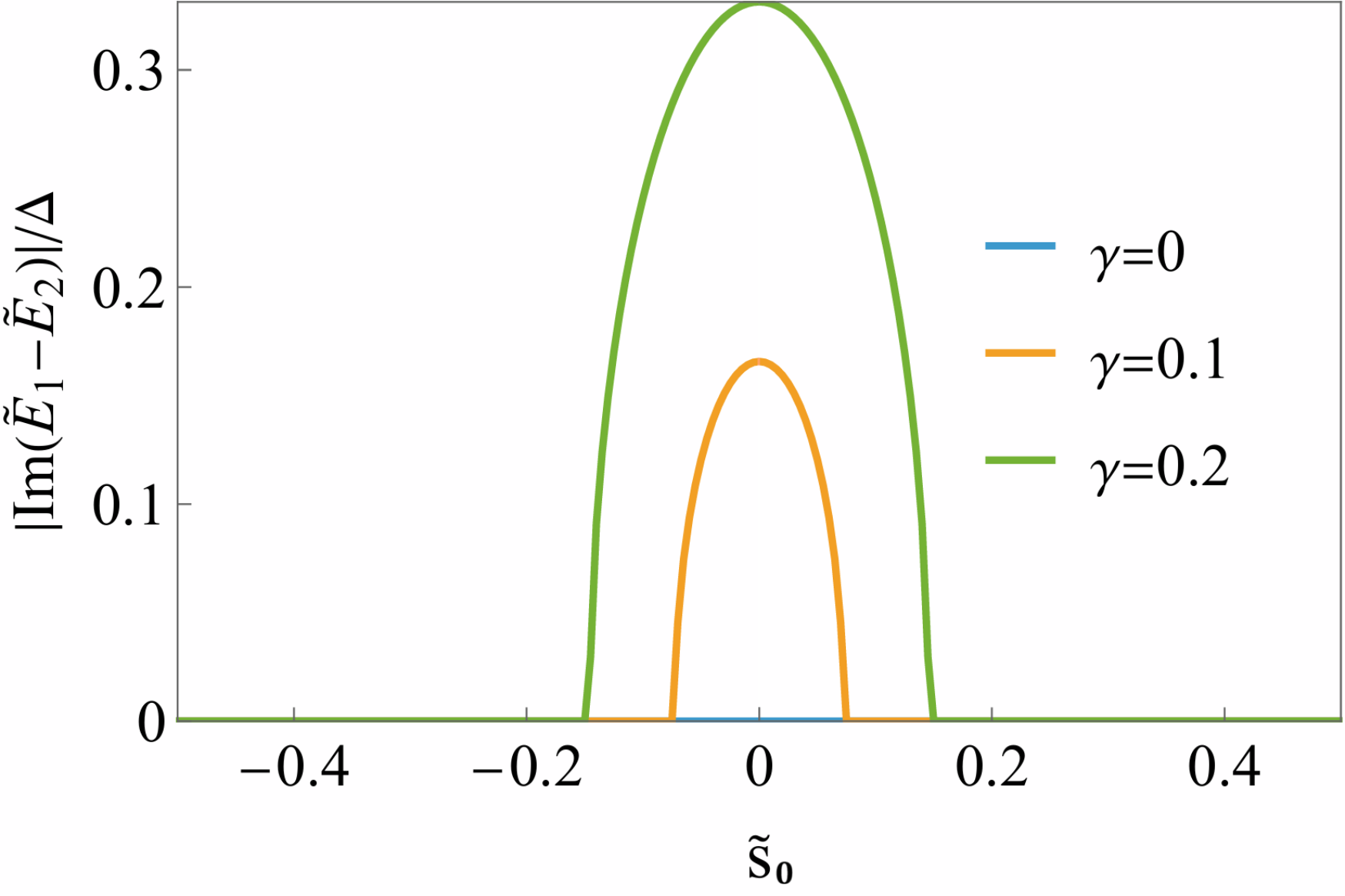}
    \put(23,60){\small\textbf{(b)}}
  \end{overpic}
\end{minipage}

\caption{The energy difference between exact eigenvalues of the effective Hamiltonian~\eqref{Heff gamma} for (a) $\BP\BT$-symmetric phase, (b) $\BP\BT$-symmetry-broken phase.}
\label{Oscillation frequency}

\end{figure*}

Varying the parameter $\tilde s_0$ we obtain that two different types of quantum dynamical behavior are realized in $\BP\BT$--symmetric two-level systems. 
Indeed, as $\tilde s_0=\pm0.2$ or $\tilde s_0=\pm 0.08$ the $\BP\BT$--symmetric two-level system falls in the $\BP\BT$--symmetry preserved regime with the energy levels taking real values (see, Fig. \ref{spectrum eff}b), and the relevant population probabilities $P_{\uparrow(\downarrow)}(t)$ demonstrate the regular quantum oscillations (see, Figs.~\ref{relative prob plots}(a,b).). 
Moreover, one can see a substantial decrease of the oscillation frequency $\omega$ and an increase in the amplitude of the oscillation as $s_0$ approaches to the \textit{EP}s (compare, the panels (a) and (b) of Fig.~\ref{relative prob plots}). This behavior is shown in Fig.~\ref{Oscillation frequency}(a) by plotting $ \hbar \omega/\Delta= |\mathrm{Re}(E_1-E_2)|/\Delta$ as a function of $\tilde s_0$ for different values of $\gamma$.

In the  $\BP\BT$--symmetry broken regime where the energy levels take complex conjugate values, the quantum oscillations are absent, and the population probability $P_{\uparrow(\downarrow)}(t)$ exponentially decreases (increases) and finally saturates on the value of $P_{\uparrow(\downarrow)}=1/2$, see Fig.~\ref{relative prob plots}(c). By increasing the gain (loss) parameter, the time of decay $\tau$ increases as well. This behavior is shown in Fig.~\ref{Oscillation frequency}(b) by plotting the parameter $\hbar/(\Delta \tau)=|\mathrm{Im}(E_1-E_2)|/\Delta$ as a function of $\tilde s_0$ for different values of $\gamma$.

To conclude this subsection we notice that similar quantum dynamics, i.e., the quantum oscillations in the $\BP\BT$-symmetry preserved regime and exponentially decaying probability of the initially prepared state in the  $\BP\BT$-symmetry broken regime, has been obtained theoretically in \cite{Wu:2019hsc} and verified in experiments with single qubits \cite{dogra2021quantum,Kazima,Wu:2019hsc}. 


\subsection{Adiabatically varying effective Hamiltonian: $s=t/T$}
In this subsection we consider the Hamiltonian \eqref{Heff gamma} slowly (adiabatically) varying in time as $s=t/T$, where $0<t<T$ and $T$ is the duration of $QAA$. This regime defines the \textit{QAA} process \cite{farhi1998analog,farhi2000quantum,king2023quantum} extended to $\BP\BT$--symmetric non-Hermitian systems. 

\begin{figure}[h]
\centering
\includegraphics[width=\columnwidth]{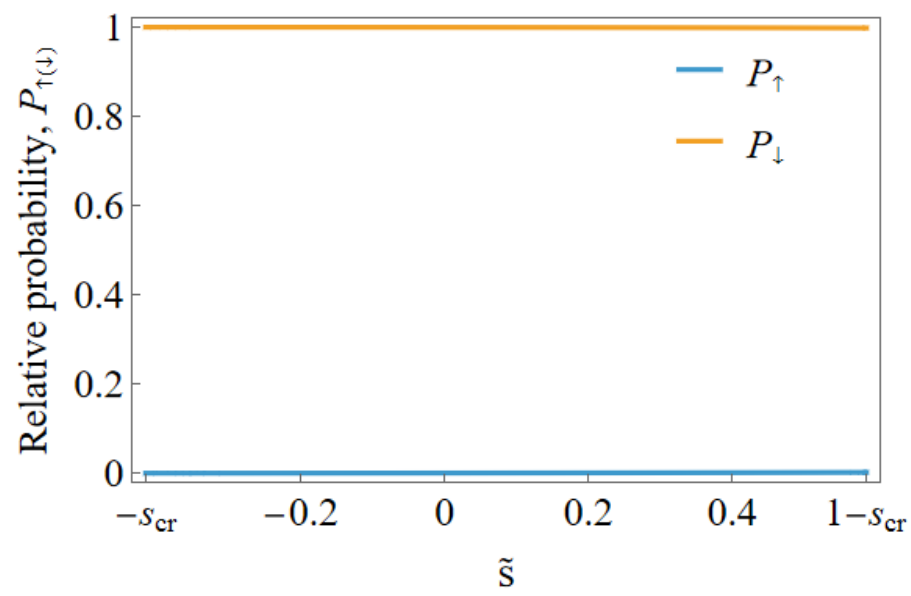}
\caption{Evolution of the relative population probability $P_{\uparrow(\downarrow)}(\tilde s)$ in the absence of gain (loss) $\gamma/\Delta=0$. The qubit's bias is absent, $\epsilon=0$ }
\label{relative prob g0}
\end{figure}

\begin{figure*}[!t]
\centering

\begin{minipage}{0.44\textwidth}
  \centering
  \begin{overpic}[width=\linewidth]{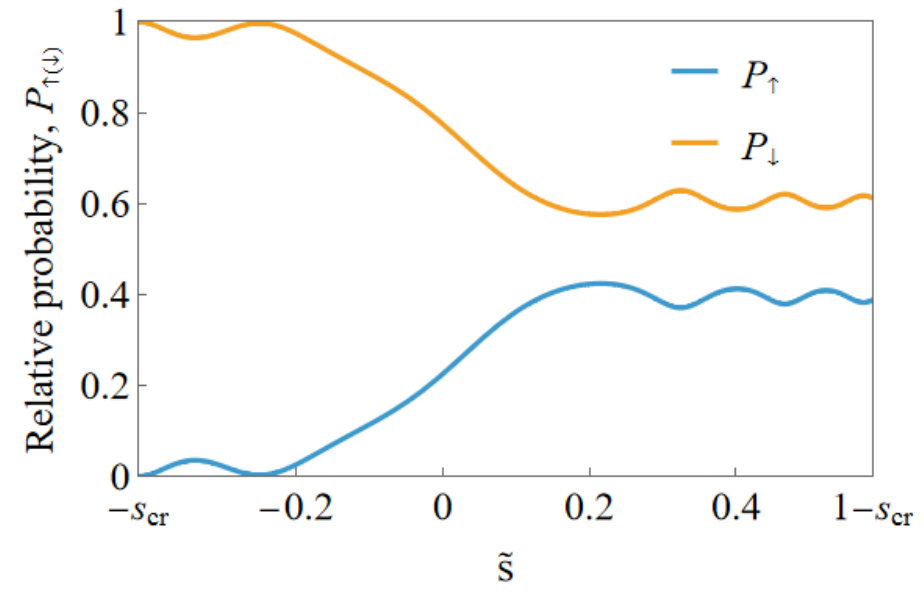}
    \put(40,57){\small\textbf{(a)} $k = 0.02$}
  \end{overpic}
\end{minipage}
\hspace{0.04\textwidth}
\begin{minipage}{0.44\textwidth}
  \centering
  \begin{overpic}[width=\linewidth]{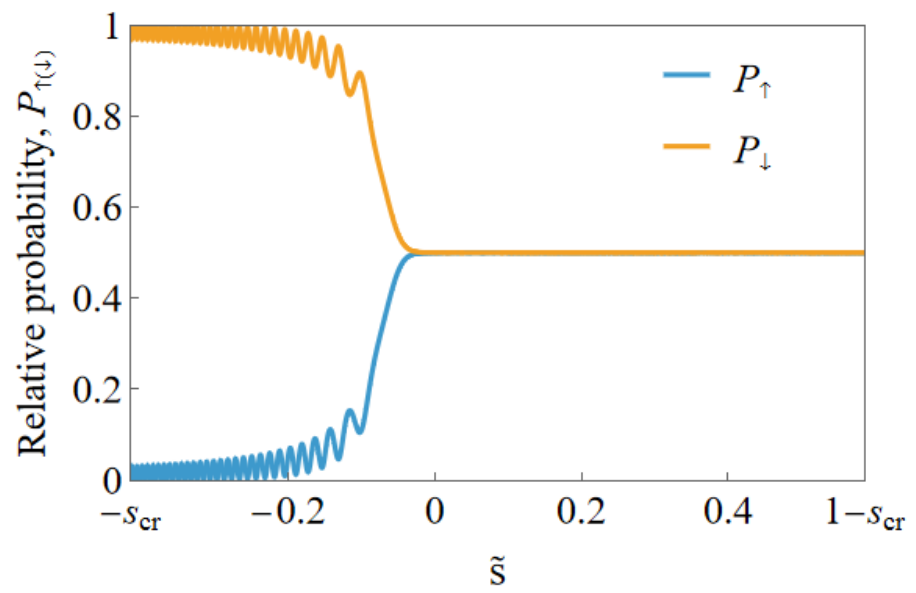}
    \put(40,57){\small\textbf{(b)} $k = 0.001$}
  \end{overpic}
\end{minipage}

\caption{Evolution of the relative population probability $P_{\uparrow (\downarrow)}$ for gain (loss) parameter $\gamma/\Delta = 0.1$: (a) $k = 0.02$, (b) $k = 0.001$. The qubit's bias is absent, $\epsilon = 0$.}
\label{relative prob eff sweep velocities}
\end{figure*}

To obtain the relevant population probabilities, $P_{\uparrow(\downarrow)}(t)$, we substitute the time $t$ to the parameter $\tilde s$ as $\tilde s=t/T-s_{cr}$ and write the Schrödinger equation \eqref{SEQq-general} in the dimensionless form as
\begin{align}
    ik\frac{d\Psi (\tilde s)}{d\tilde s}=\frac{\hat H_{eff}(\tilde s)}{\Delta}\Psi (\tilde s),
\label{SEQq}
\end{align}
where the dimensionless parameter $k=\hbar/(\Delta T)$ determines a  "speed" of \textit{QAA}. Choosing the initial state as in \eqref{psi init eff} we numerically calculate the wave functions $\psi_{\uparrow (\downarrow)}(\tilde s) $ and the 
relevant population probabilities $P_{\uparrow(\downarrow)}(\tilde s)$ for different values of the parameter $k$.  The typical dependencies of $P_{\uparrow(\downarrow)}(\tilde s)$ are presented in
Figs. \ref{relative prob g0} ($\gamma=0$) and \ref{relative prob eff sweep velocities}  ($\gamma \neq 0$). 

In the regime of $\epsilon/\Delta < 1$, Figs.~\ref{spectrum eff}(a,c) 
demonstrate that in the absence of gain (loss), $\gamma/\Delta = 0$,  the two energy levels intersect. In this case the effective  Hamiltonian~(\ref{Heff gamma}) becomes the Hermitian one, and its adiabatic time evolution corresponds to the Landau–Zener–St\"uckelberg (\textit{LZS}) type scenario \cite{Landau1932,Zener1932,Stueckelberg1932} without an avoided crossing. 
As the system is initially prepared in the ground state and the control parameter $\tilde s$ is varied through the crossing point, the state adiabatically follows the continuous branch of instantaneous eigenvalues. However, because of the exact crossing, eventually this branch brings the system to the excited state. Such  \textit{LZS} tunneling is shown in Fig.~\ref{relative prob g0}:  the system initially prepared in the state $|\downarrow\rangle$ at the end of the \textit{QAA} process remains in the same state, but which corresponds to the excited state at $s=1$. 
Therefore, even in the limit of $T$ going to infinity the system ultimately arrives on the excited state, and \textit{QAA} process becomes useless. 

As the gain/loss parameter $\gamma$ is switched on the former crossing points no longer remain simple degeneracies. Instead, they transform into regions of $\BP\BT$--symmetry broken states, where the two eigenvalues form a complex-conjugate pair, and the $\BP\BT$--symmetric non-Hermitian \textit{LZS} tunneling \cite{PhysRevA.100.062514,kivela2024quantum,pan2024nonadiabatic,erdamar2026exploring} occurs. 

In Fig.~\ref{relative prob eff sweep velocities} the typical  evolutions of the relative population probabilities of the states $|\uparrow \rangle$ ($|\downarrow \rangle$)
for two values of $k=\hbar/(\Delta T)$: $k=0.02$ and $k=0.001$, are shown. For a fast $QAA$ (moderate values of $k < 1$), see Fig.~\ref{relative prob eff sweep velocities}(a), the system initially prepared in the state $|\downarrow\rangle$ after going through \textit{EP}, displays small amplitude quantum beats, and eventually arrives on the superposition of states in which the relative population probability of the excited state ($P_\downarrow$) is slightly higher than the one of the ground 
state ($P_\uparrow$). For a slow $QAA$ ($k\ll 1$) presented in Fig.~\ref{relative prob eff sweep velocities}(b), quantum beats of a substantial amplitude appear before going through \textit{EP} and eventually the system arrives on the equal superposition of states in which the relative population probability of being either in $|\downarrow\rangle$ or in $|\uparrow\rangle$ is $0.5$.

\begin{figure*}[!t]
\centering

\begin{minipage}{0.44\textwidth}
  \centering
  \begin{overpic}[width=\linewidth]{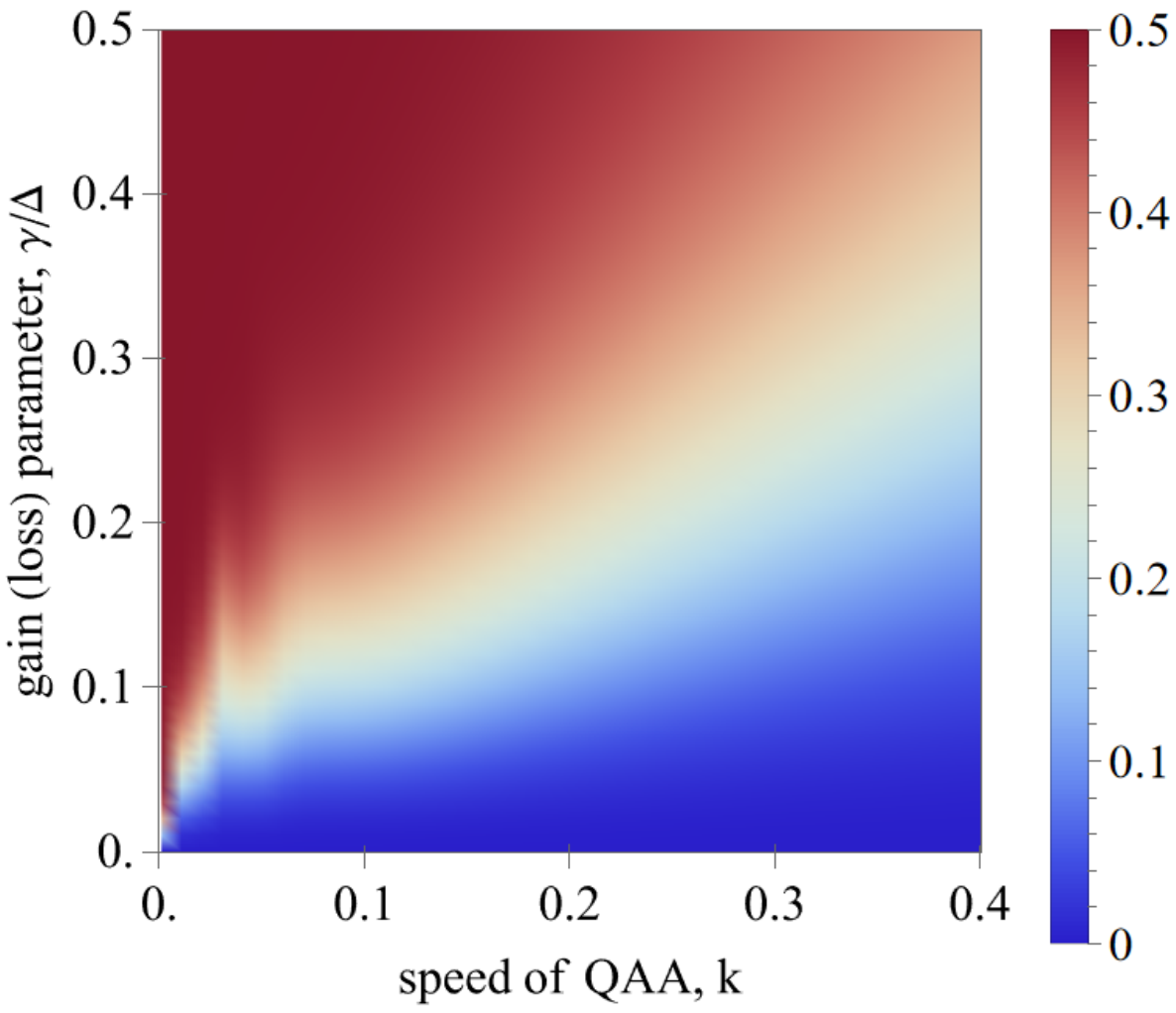}
    \put(16,76){\small\textbf{(a)}}
  \end{overpic}
\end{minipage}
\hspace{0.04\textwidth}
\begin{minipage}{0.44\textwidth}
  \centering
  \begin{overpic}[width=\linewidth]{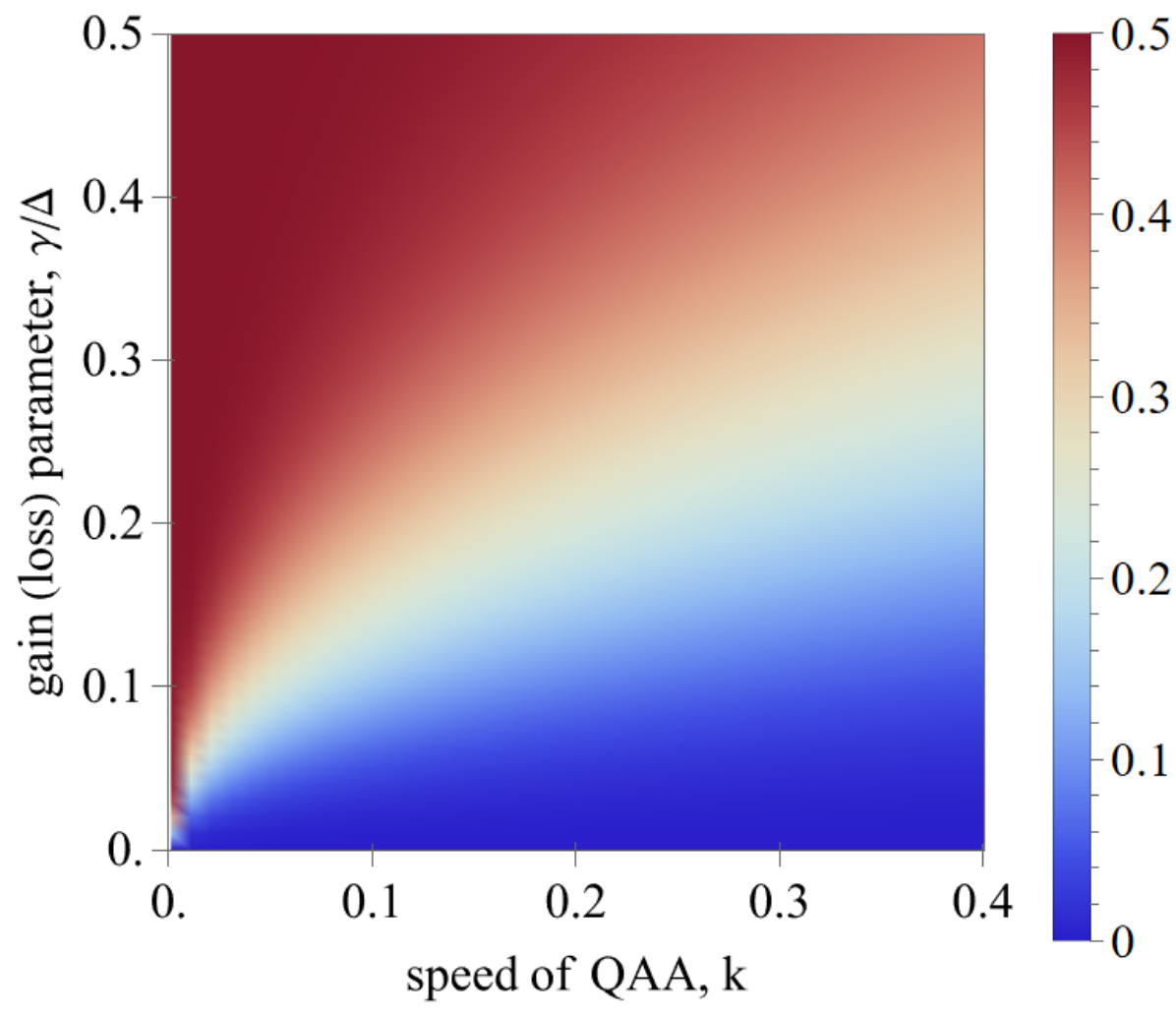}
    \put(16,76){\small\textbf{(b)}}
  \end{overpic}
\end{minipage}
\caption{Numerically (a) and analytically (b) calculated dependencies of the population probability $P_\uparrow$ on the gain (loss) parameter $\gamma$ and the speed of \textit{QAA}, demonstrating the \textit{LZS} tunneling in $\BP\BT$--symmetric non-Hermitian two-level systems, are presented as color plots. The qubit's bias is absent, $\epsilon = 0$.}
\label{model analytical e0}
\end{figure*}
%



\begin{figure*}[!t]
\centering

\begin{minipage}{0.44\textwidth}
  \centering
  \begin{overpic}[width=\linewidth]{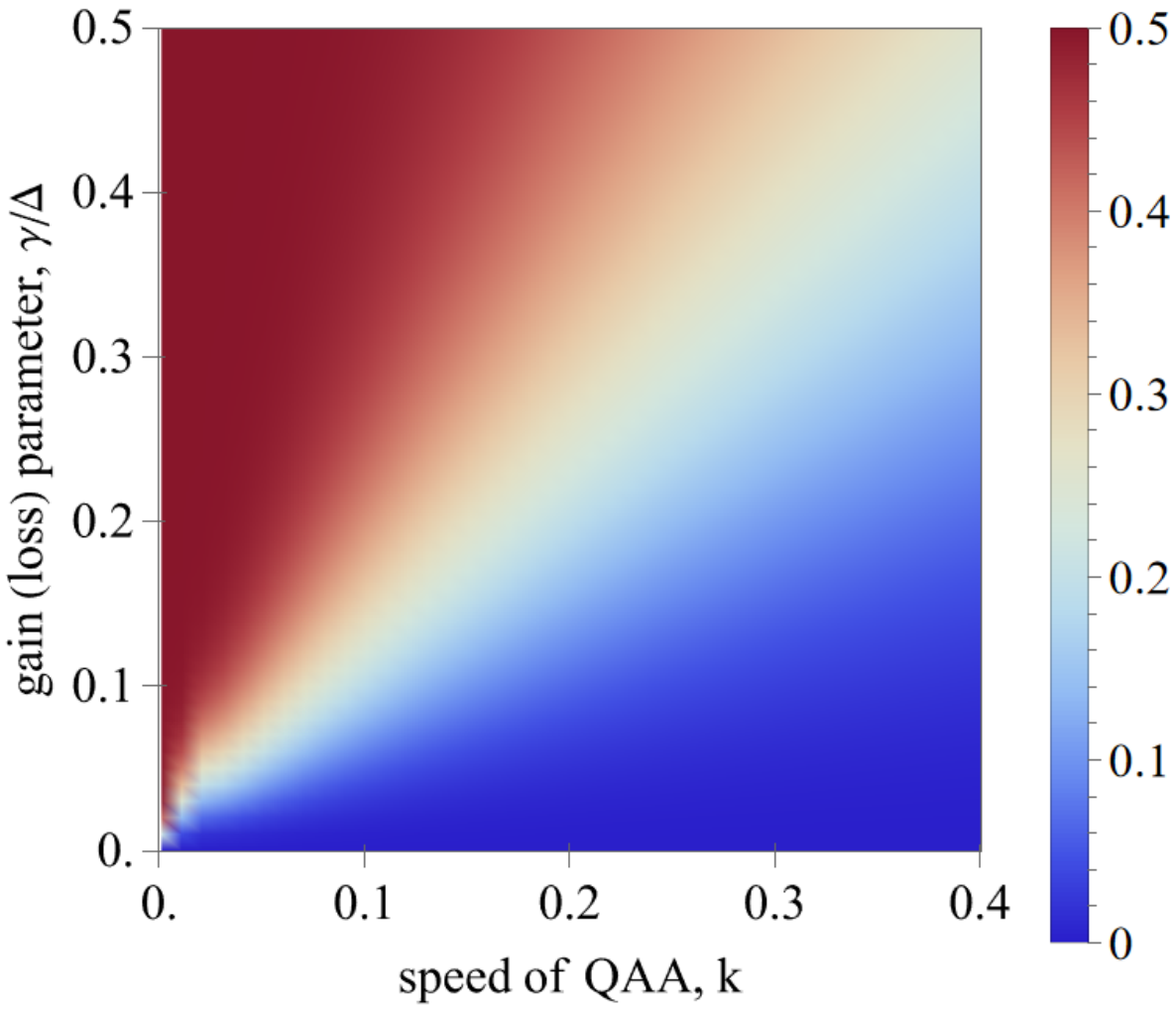}
    \put(16,76){\small\textbf{(a)}}
  \end{overpic}
\end{minipage}
\hspace{0.04\textwidth}
\begin{minipage}{0.44\textwidth}
  \centering
  \begin{overpic}[width=\linewidth]{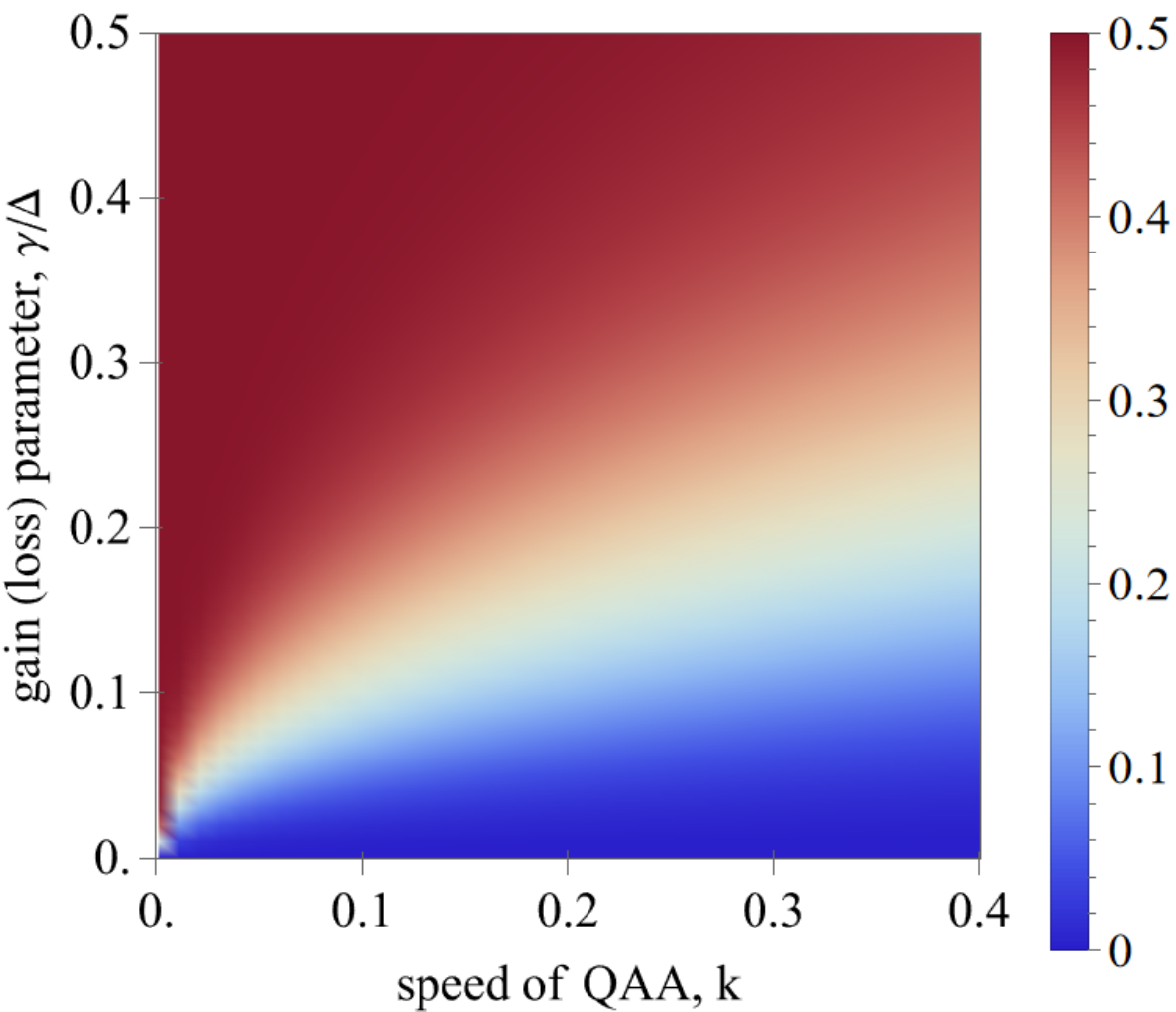}
    \put(16,76){\small\textbf{(b)}}
  \end{overpic}
\end{minipage}
\caption{Numerically (a) and analytically (b) calculated dependencies of the population probability $P_\uparrow$ on the gain (loss) parameter $\gamma$ and the speed of \textit{QAA}, demonstrating the \textit{LZS} tunneling in $\BP\BT$--symmetric non-Hermitian two-level systems, are presented as color plots. The qubit's bias was chosen as $\epsilon/\Delta=0.9$.}
\label{model analytical e09}
\end{figure*}

To explore the possibility of improving the \textit{QAA }process using $\BP\BT$--symmetric non-Hermitian qubits networks we numerically calculate the probability to find the effective $\BP\BT$--symmetric two levels system in the ground state of the Hermitian part of the  Hamiltonian (\ref{Heff gamma}) at time $t=T$, $P_{\uparrow}(s=1)$ for different gain (loss) parameter $\gamma$ and $QAA$ duration $T$. These results are shown  as the color plots in Fig.~\ref{model analytical e0}(a) for the unbiased qubits, $\epsilon=0$, and in Fig.~\ref{model analytical e09}(a) for the biased qubits,  $\epsilon/\Delta=0.9$.

Probability to be in the ground state of the Hermitian part of the  Hamiltonian (\ref{Heff gamma}) \textit{determined by} the \textit{LZS} tunneling in $\BP\BT$--symmetric non-Hermitian two-level systems, can be found analytically by adapting the method elaborated in Refs.~\cite{kayanuma1984,kayanuma1998,grifoni1998driven} and previously used for an analysis of various ac driven two-level systems. This probability $P(\gamma,k,\epsilon)$ is explicitly expressed as (see details in Appendix ~\ref{section:LZS derivation})
\begin{equation}
    P(\gamma,k,\epsilon)=\frac{\exp \left[\frac{2|\ell|^2 \pi}{(g-w)k}\right]-1}{2\exp \left[\frac{2|\ell|^2 \pi}{(g-w)k}\right]-1}
\label{analytical eff eq}
\end{equation}

The dependencies of $P(\gamma,k,\epsilon)$ on the parameters $\gamma$ and $k$ are shown as the color plots in Figs.~\ref{model analytical e0}(b) for unbiased qubits, and ~\ref{model analytical e09}(b) for biased qubits. Comparing the Figs.~\ref{model analytical e0}-\ref{model analytical e09}(a) and \ref{model analytical e0}-\ref{model analytical e09}(b) one can see a good agreement between direct numerical calculations and analytically derived Eq. \eqref{analytical eff eq}.  Since Eq. \eqref{analytical eff eq} is valid under the condition $\tilde s_f \sqrt{(g-w)/k\Delta} \gg 1$ (see Appendix for details) one can also notice that the agreement becomes definitely worse for the biased qubits  and as $k$ is not too small (see, Fig.~\ref{model analytical e09}).

\section{Quantum annealing with two interacting $\BP\BT$--symmetric qubits}\label{section:QAA original problem}

Here, we turn back to the original Hamiltonian ~\eqref{hamiltonian full 2q} and provide a numerical study of the $QAA$ process for two interacting $\BP\BT$--symmetric non-Hermitian qubits.  For that we use the following method: firstly, we determine the exact eigenvalues $E_{1-4}(s)$ and eigenvectors $\vec \psi_{1-4}(s)$ of the \textit{Hermitian part} of the Hamiltonian~\eqref{hamiltonian full 2q}, i.e., \eqref{hamilton matrix} for $\gamma=0$, for the initial ($s=0$) and final ($s=1$) values of the parameter $s$.  The minimal eigenvalue $E_1(s)$ defines the instantaneous ground state of the Hermitian Hamiltonian~\eqref{hamiltonian full 2q} with $\gamma=0$, 
whose eigenvector $\vec \psi_1(s)$ corresponds to the initial state $\vec \psi_1(0)$ at $s=0$ and to the target ground state $\vec \psi_1(1)$ at $s=1$ of the successful $QAA$. The eigenvectors $\vec \psi_{1-4}(1)$ at $s=1$ will serve as basis vectors in the expansion of the wave function $\Psi (s=1)$. Secondly, we numerically solve  the Schrödinger equation written, similarly to Eq.~(\ref{SEQq}), in the dimensionless form as:
\begin{align}
    ik\frac{d\Psi ( s)}{ds}=\frac{\hat H(s)}{\Delta}\Psi ( s)
\label{SEQq-complete}
\end{align}
with $k=\hbar/(\Delta T)$. This equation is accompanied by the initial condition chosen as $\Psi(0)=\vec \psi_1(s=0)$. Finally, we expand the resulting wave function $\Psi(s=1)$ into basis eigenvectors, $ \vec \psi_{1-4}(s=1)$, as
\begin{align}
    \Psi(s=1)=\sum_{i=1}^{4} a_i \vec \psi_i(s=1),
\end{align}
where the overlap coefficients $a_i$ are given by
\begin{align}
    a_i = \langle \vec \psi_i (s=1), \Psi (s=1) \rangle = \vec \psi_i^{\dagger}\Psi.
\label{coeff eq}
\end{align}
Using the calculated $a_i$ we determine the probability of $P_{gr}$ to obtain the system in the ground state of Hermitian Hamiltonian $\hat H_f$ \eqref{final hamiltonian} as
\begin{equation}
    P_{gr}=\frac{|a_{1}|^2}{\sum_{i=1}^4|a_i|^2}.
\end{equation}

\begin{figure*}[!t]
\centering

\begin{minipage}{0.44\textwidth}
  \centering
  \begin{overpic}[width=\linewidth]{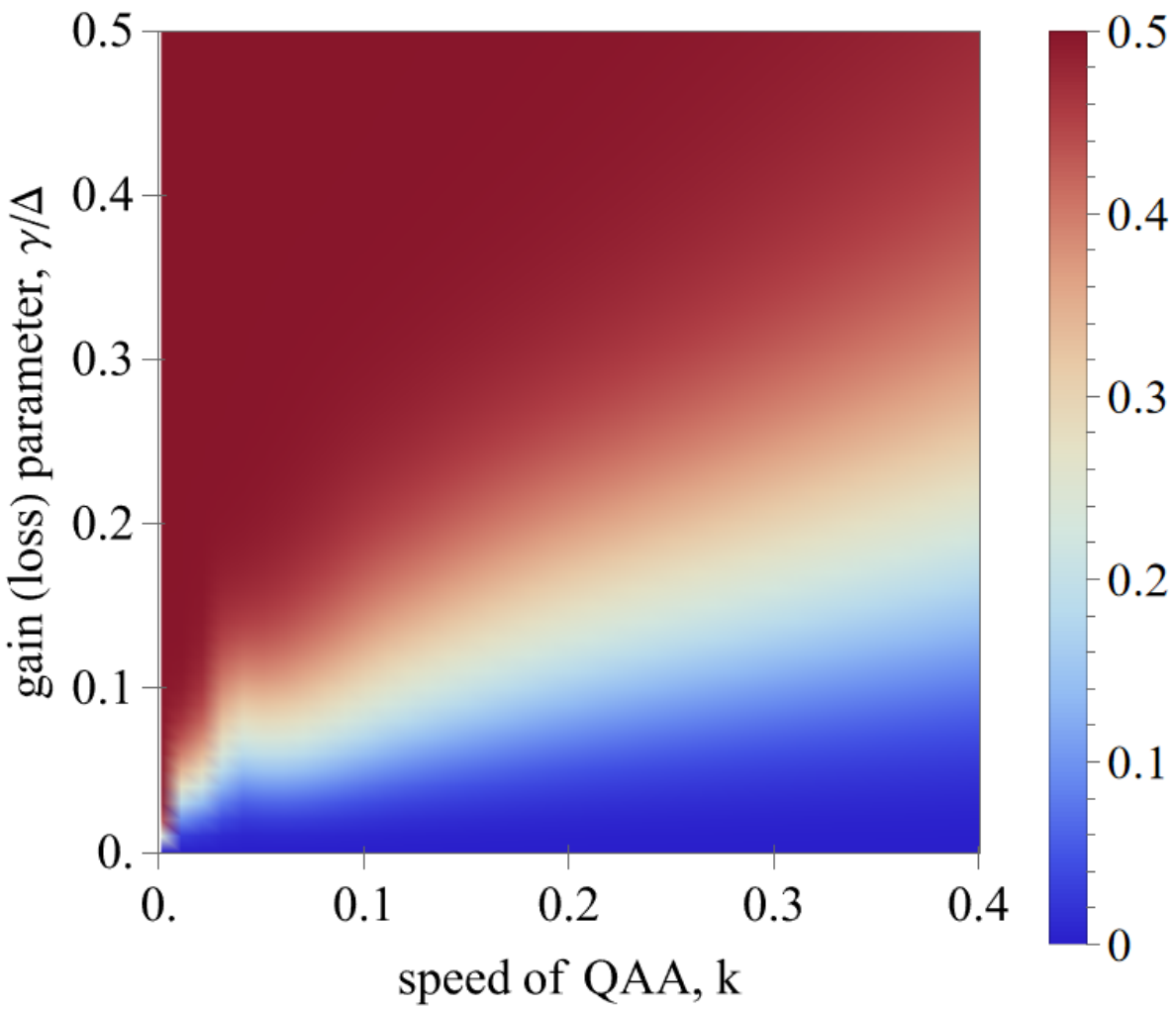}
    \put(16,76){\small\textbf{(a)}}
  \end{overpic}
\end{minipage}
\hspace{0.04\textwidth}
\begin{minipage}{0.44\textwidth}
  \centering
  \begin{overpic}[width=\linewidth]{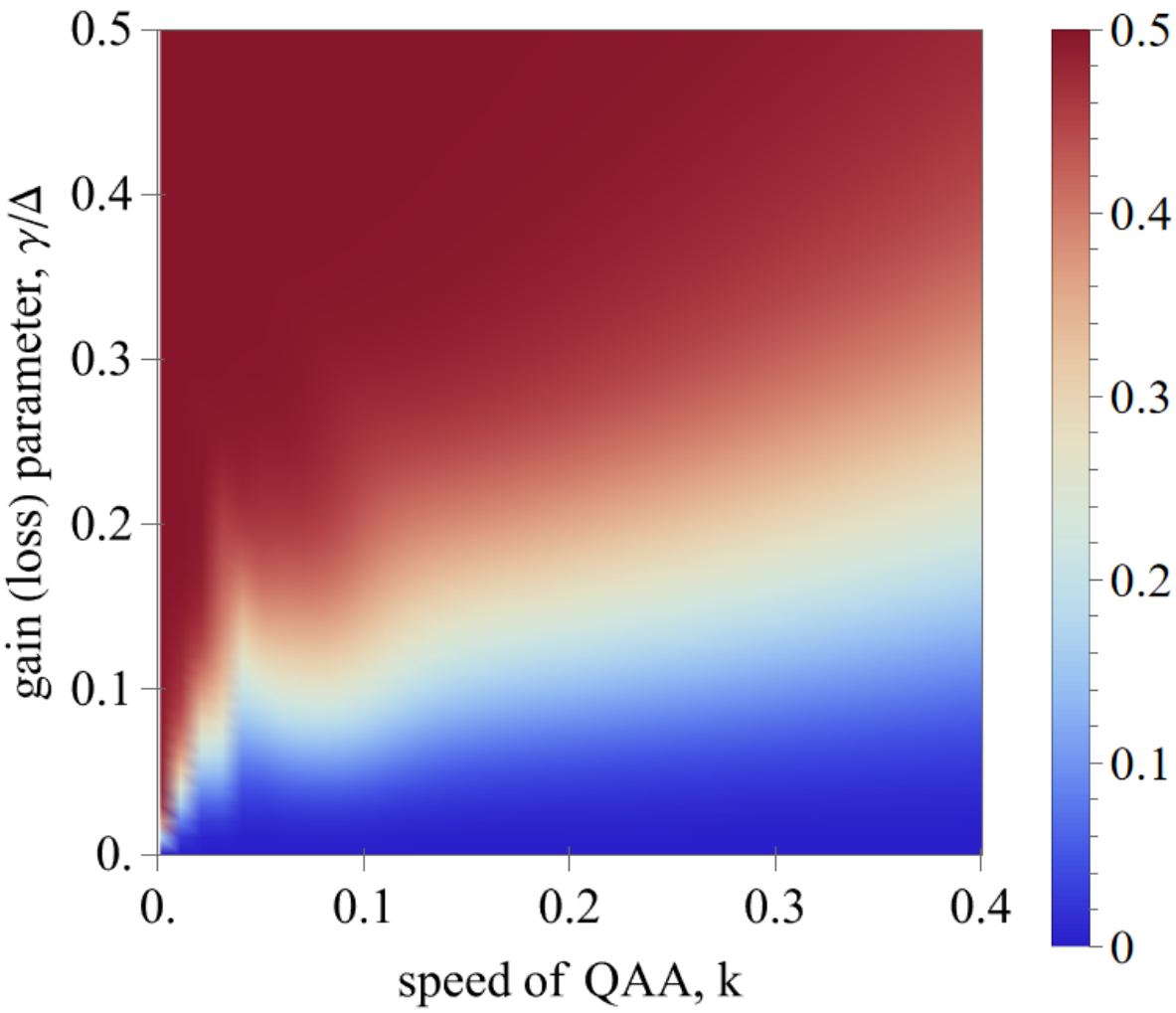}
    \put(16,76){\small\textbf{(b)}}
  \end{overpic}
\end{minipage}
\caption{The dependencies of the probability to be in the ground state of the Hamiltonian  $\hat H_f$ \eqref{final hamiltonian} for two interacting $\BP\BT$--symmetric qubits on 
the gain (loss) parameter $\gamma$ and the speed of \textit{QAA } process $k$. The color plots are presented for: a) unbiased qubits; b) biased qubits ($\epsilon=0.9$). The interaction strength was chosen as $g/\Delta=1$}
\label{color plot full}
\end{figure*}

The dependencies of $P_{gr}$ on the gain (loss) parameter $\gamma$ and the duration $T$ of $\BP\BT$--symmetric non-Hermitian \textit{QAA}, demonstrating the possibility of a successful $QAA$, are presented in Fig.~\ref{color plot full}(a,b) for unbiased ($\epsilon/\Delta=0$) and biased ($\epsilon/\Delta=0.9$) interacting qubits. 

\section{Conclusion and Discussion}\label{section:Conclusions}

We have analyzed the quantum dynamics of two interacting qubits governed by a time-dependent $XX$-type $\mathcal{PT}$-symmetric non-Hermitian Hamiltonian and assessed its performance in adiabatic quantum annealing. By examining the instantaneous spectrum, we identified regimes of unbroken and broken $\mathcal{PT}$ symmetry and constructed an effective two-level description capturing the essential low-energy behavior.

In the time-independent setting, the system exhibits two distinct dynamical regimes. In the $\mathcal{PT}$-symmetric phase, the population shows oscillatory behavior with a frequency that decreases near the exceptional point, accompanied by enhanced amplitude. In the symmetry-broken phase, oscillations vanish and the dynamics becomes exponential, leading to saturation at equal population.

For the annealing process, we find that the Hermitian limit fails to reliably reach the ground state, even for slow driving. In contrast, introducing a $\mathcal{PT}$-symmetric non-Hermitian term significantly improves performance: the ground-state population can be enhanced and reaches a maximal value of 1/2 at moderate annealing speeds.

We also identify characteristic features of non-Hermitian annealing dynamics. For fast driving, oscillations appear after crossing the symmetry-broken region, while for slow annealing, pronounced quantum beating emerges before the exceptional point, followed by saturation at equal population. Numerical and analytical results for the effective model show strong agreement, and the full two-qubit system displays qualitatively similar behavior.

Overall, our results demonstrate that $\mathcal{PT}$-symmetric non-Hermitian terms can enhance the efficiency of adiabatic quantum annealing. Extending these findings to many-qubit systems, clarifying the role of bias, and pursuing experimental validation are important directions for future work.

\begin{acknowledgments}
 We thank Sergej Flach and Jung-Wan Ryu for fruitful discussions. M.V.F acknowledge the hospitality and the partial financial support of the Center for Theoretical Physics of Complex Systems (PCS), Institute for Basic Science (IBS), Republic of Korea and the International Institute of Physics (IIP), Natal Brazil.
\end{acknowledgments}

\appendix

\section{\textit{LZS} tunneling in $\BP\BT$-symmetric non-Hermitian two-level systems}\label{section:LZS derivation}
Here, we present detailed derivation of the population probability $P(\gamma, k, \epsilon)$ to be in the ground state of the Hermitian part of the Hamiltonian \eqref{Heff gamma}. The $\BP\BT$--symmetric non-Hermitian Hamiltonian $\hat H_{eff}$ of a two-level system is written as
\begin{equation}
\hat H_{eff}=\left(
\begin{array}{cc}
 H_{11}[\tilde s] & i\ell  \\
 i\ell  & H_{22}[\tilde s] \\
\end{array}
\right),
\label{Heff}
\end{equation}
where $H_{11}=-g\tilde s$ and $H_{22}=-w\tilde s$. The \textit{LZS} tunneling is determined by the time-dependent Hamiltonian \eqref{Heff} with $\tilde s$ varies linearly from $\tilde s=\tilde s_{in}$ to $\tilde s=\tilde s_f$. The wave function $\Psi(\tilde s )$ is satisfied to the dimensionless Schr\"odinger equation \eqref{SEQq}. 

Next, extending  the region of $\tilde s$ up to  $\tilde s=-\infty$ and using the method elaborated in \cite{kayanuma1984,kayanuma1998,grifoni1998driven} we present the wave function $\psi_\downarrow (\tilde s_f)$ in the following form:

\begin{align}
\psi_\downarrow (\tilde s_f)= & \sum_{n=0}^{\infty}\left (\frac{\ell}{k \Delta} \right)^{2n} \int_{-\infty}^{\tilde s_f} d \tilde s_{2n} \int_{-\infty}^{s_{2n}} d \tilde s_{2n-1}...\int_{-\infty}^{\tilde s_2} d \tilde s_{1} \cdot \nonumber \\ 
& \cdot \exp \left \{-i\frac{1}{k\Delta}\sum_{m=0}^{n} \mathcal{F}(m,n; \tilde s_0...\tilde s_{2n+1}) \right \},
\label{Wave function 1}
\end{align}
where the functional $\mathcal{F}$ is expressed as
\begin{align}
\mathcal{F}(m,n; \tilde s_0...\tilde s_{2n+1})= &  \int_{\tilde s_{2m}}^{\tilde s_{2m+1}} H_{11}(x_m) dx_m  + \nonumber \\
&+\int_{\tilde s_{2m+1}}^{\tilde s_{2m+2}}H_{22}(x_m)  dx_m \}.
\label{Wave function 2}
\end{align}
Here, $\tilde s_0=-\infty$ and $\tilde s_{2n+1}=\tilde s_f$. 
Calculating the integrals in \eqref{Wave function 2} we obtain
\begin{align}
\psi_\downarrow (\tilde s_f)= & \sum_{n=0}^{\infty}\left (\frac{\ell}{k \Delta} \right)^{2n} \int_{-\infty}^{\tilde s_f} d \tilde s_{2n} \int_{-\infty}^{s_{2n}} d \tilde s_{2n-1}...\int_{-\infty}^{\tilde s_2} d \tilde s_{1} \cdot \nonumber \\ 
& \cdot \exp \left \{-i\frac{(g-w)}{2k\Delta}\sum_{m=0}^{2n} (-1)^m \tilde s_{m}^2\right \}.
\label{Wave function 3}
\end{align}
The $\tilde s$-ordered integrals in \eqref{Wave function 3} can be carried out with the substitutions \cite{kayanuma1984}:
\begin{align}
x_1= & \tilde s_1 \\
x_p= &\tilde s_1+\sum_{m=0}^{p-1} \tilde s_{2m+1}-\tilde s_{2m}, ~2\leq p \leq n \\
y_p= & \tilde s_{2p}-\tilde s_{2p-1}, ~1\leq p \leq n. 
\label{Substitution}
\end{align}
In new variables \eqref{Wave function 3} reads as
\begin{align}
\psi_\downarrow (\tilde s_f)= & \sum_{n=0}^{\infty}\left (\frac{\ell}{k \Delta} \right)^{2n} \mathcal{A}_n
\label{Wave function 4}
\end{align}
with 
\begin{align}
\mathcal{A}_n=& \int_{-\infty}^{\tilde s_f} dx_1 \int_{x_1}^{\tilde s_f} dx_2...\int_{x_{n-1}}^{\tilde s_f} dx_n  \cdot  \nonumber  \\
& \cdot \int_{0}^{\infty}  dy_1 \int_{0}^{\infty} dy_2...\int_{0}^{\infty} dy_n \cdot \nonumber \\
& \cdot \exp \left \{ i \frac{g-w}{2k\Delta} \left [ \left (\sum_{p=1}^n y_p \right)^2+2\sum_{p=1}^n x_py_p \right ] \right \}.
\label{A-integrals}
\end{align}
Since the integrals are unchanged by arbitrary permutations of the variables $\{x_n\}$ we rewrite (\ref{A-integrals}) as
\begin{align}
\mathcal{A}_n=& \frac{1}{n !} \prod_{p=1}^n\int_{-\infty}^{\tilde s_f} dx_p \int_{0}^{\infty}  dy_p \cdot \nonumber \\
& \cdot \exp \left \{ i \frac{g-w}{2k\Delta} \left [ \left (\sum_{p=1}^n y_p \right)^2+2\sum_{p=1}^n x_py_p \right ] \right \}.
\label{A-integrals-2}
\end{align}
The $\mathcal{A}_n$ can be calculated in the compact form by making use of the following representation:
\begin{align}
\mathcal{A}_n=& \frac{1}{n !}\int_{-\infty}^{\infty} \int_{-\infty}^{\infty}\frac{d\xi d\eta}{2\pi}\prod_{p=1}^n\int_{-\infty}^{\tilde s_f} dx_p \int_{0}^{\infty}  dy_p \cdot \nonumber \\
& \cdot \exp \left \{ i\xi \eta-i\xi y_p + i \frac{g-w}{2k\Delta} \left [ \eta^2+2\sum_{p=1}^n x_py_p \right ] \right \}.
\label{A-integrals-3} 
\end{align}
Calculating the integral over $\eta$ we obtain
\begin{align}
\mathcal{A}_n=& \frac{1}{n !}\sqrt{\frac{2k\Delta}{g-w}}\int_{-\infty}^{\infty} \frac{d\xi}{2\sqrt{\pi}} \exp \left [-i \frac{k\Delta}{2(g-w)}\xi^2 \right ] \cdot \nonumber \\
&\cdot \prod_{p=1}^n\int_{-\infty}^{\tilde s_f} dx_p \int_{0}^{\infty}  dy_p \exp \left \{ -i\xi y_p + i \frac{g-w}{k\Delta} \sum_{p=1}^n x_py_p \right \}.
\label{A-integrals-4} 
\end{align}
Combining \eqref{Wave function 4} and \eqref{A-integrals-4} we obtain
\begin{align}
\psi_\downarrow (\tilde s_f)= & \sqrt{\frac{2k\Delta}{g-w}}\int_{-\infty}^{\infty} \frac{d\xi}{2\sqrt{\pi}} \exp \left [-i \frac{k\Delta}{2(g-w)}\xi^2 \right ] \cdot \nonumber \\
& \cdot \exp \left \{ \frac{\ell^2}{k^2 \Delta^2} \mathcal{I}(\xi) \right \},
\label{Wave function-Compact}
\end{align}
where $\mathcal{I}(\xi)$ is expressed as 
\begin{align}
\mathcal{I}(\xi)=& \int_{-\infty}^{\tilde s_f} dx \int_{0}^{\infty }  dy \exp \left \{ -i\xi y + i \frac{g-w}{k\Delta} xy \right \}.
\label{A-integrals-41} 
\end{align}
Calculating the integral over $y$ we obtain
\begin{align}
\mathcal{I}(\xi)=& \int_{-\infty}^{\tilde s_f} dx \frac{1}{i\left (\xi-\frac{g-w}{k\Delta} x-i\lambda \right)},
\label{A-integrals-5} 
\end{align}
where $\lambda$ is infinitesimal positive constant.  
As the condition of $\tilde s_f \sqrt{(g-w)/k\Delta} \gg 1$ is valid the integrals over  $x$ and $\xi$ can be easily calculated, and we obtain
\begin{align}
|\psi_\downarrow (\tilde s_f)|^2= \exp \left \{ \frac{2 \pi \ell^2}{(g-w)k\Delta} \right\}.
\label{Wave function_down}
\end{align}
Similarly to that we yeld  the $|\psi_\uparrow (\tilde s_f)|^2$ in the following form \cite{kayanuma1998}:
\begin{align}
|\psi_\uparrow (\tilde s_f)|^2= -\sum_{n=1}^\infty \left (\frac{\ell}{k\Delta} \right)^{2n} \mathcal{L}^{(n)}
\label{probability_1}
\end{align}
and
\begin{align}
 \mathcal{L}^{(n)}=\sum_{m=0}^{n} \mathcal{A}_m \mathcal{A}^\star_{n-m}.
\label{Ell}
\end{align}
By making use of Eqs.~(\ref{A-integrals-2})-(\ref{A-integrals-3}) we obtain explicitly 
\begin{align}
|\psi_\uparrow (\tilde s_f)|^2= \exp \left \{ \frac{2 \pi \ell^2}{(g-w)k\Delta} \right\}-1.
\label{Wave function_up}
\end{align}
Using  (\ref{Probability def}) finally we arrive on the expression \eqref{analytical eff eq} for the probability $P(\gamma, k, \epsilon)$.
\bibliography{literatur}
\end{document}